\documentclass[%
 reprint,
 superscriptaddress,
 amsmath,amssymb,
 aps,
 prl,
floatfix,
]{revtex4-2}

\usepackage{orcidlink}
\usepackage{graphicx}% Include figure files
\usepackage{dcolumn}% Align table columns on decimal point
\usepackage{booktabs}
\usepackage{makecell}
\usepackage{bm}% bold math
\usepackage{lipsum}
\usepackage{hyperref}
\hypersetup{colorlinks=true,allcolors=blue}
\usepackage[normalem]{ulem}
\usepackage{lineno}
\usepackage{float} 
\usepackage{amsmath}

\usepackage[separate-uncertainty]{siunitx}
\DeclareSIUnit\years{yrs}
\DeclareSIUnit\pe{PE}
\DeclareSIUnit\years{years}
\DeclareSIUnit\photoelectron{PE}
\DeclareSIUnit\inch{"}
\usepackage{printlen}
\usepackage{wrapfig}
\usepackage[dvipsnames]{xcolor}

\usepackage[rightcaption]{sidecap}
\sidecaptionvpos{figure}{c}

\newif\ifincludesupplement
\includesupplementtrue % <-- turn ON to include supplement
\newcommand{\ie}{{\it i.e.}}

\newcommand{\bi}{\begin{itemize}}
	\newcommand{\ei}{\end{itemize}}

\begin{document}

\title{First Constraints on Long-Range Neutrino Interactions using IceCube DeepCore}

%%%%%

\author{Gopal Garg\,\orcidlink{0009-0003-3134-2089}}
\affiliation{Institute of Physics, Sachivalaya Marg, Sainik School Post, Bhubaneswar 751005, India}
\affiliation{Department of Physics, Aligarh Muslim University, Aligarh 202002, India}
%
%%%%%
%
\author{J Krishnamoorthi\,\orcidlink{0009-0006-1352-2248}}
\affiliation{Institute of Physics, Sachivalaya Marg, Sainik School Post, Bhubaneswar 751005, India}
\affiliation{Department of Physics, Aligarh Muslim University, Aligarh 202002, India}
%
%%%%%
%
\author{Anil Kumar\,\orcidlink{0000-0002-8367-8401}}
\affiliation{Institute of Physics, Sachivalaya Marg, Sainik School Post, Bhubaneswar 751005, India}
%
%
%%%%%
\author{Sanjib Kumar Agarwalla\,\orcidlink{0000-0002-9714-8866}}
\affiliation{Institute of Physics, Sachivalaya Marg, Sainik School Post, Bhubaneswar 751005, India}
\affiliation{Homi Bhabha National Institute, Training School Complex, Anushakti Nagar, Mumbai 400094, India\\
{\tt gi8820@myamu.ac.in, krishnamoorthi.j@iopb.res.in, anil.k@iopb.res.in, sanjib@iopb.res.in} \smallskip}
\date{\today}

\begin{abstract}
We present the first search for new flavor-dependent long-range interactions (LRI) of neutrinos using publicly available 8 years of high-purity $\nu_\mu$ CC data from IceCube DeepCore. These interactions are mediated by ultra-light gauge bosons with masses below $10^{-10}$ eV, which can arise due to a new lepton-number gauge symmetry, such as $L_e - L_\mu$ or $L_e - L_\tau$. These long-range interactions induce matter potential between neutrinos and abundant electrons present in distant astrophysical sources. These LRI potentials could modify neutrino oscillation probabilities. By probing the effects of LRI on atmospheric neutrino oscillations at IceCube DeepCore, we place world-leading constraints on the coupling strength of these  interactions.

\end{abstract}

\maketitle

\noindent \textbf{Introduction.---} The discovery of neutrino oscillations~\cite{Super-Kamiokande:1998kpq,SNO:2002tuh, KamLAND:2002uet,DayaBay:2012fng,IceCube:2014flw} established that neutrinos possess nonzero masses and undergo flavor mixing, highlighting the need for an extension of the Standard Model (SM) of particle physics. This makes neutrinos a unique probe to search for potential signatures of physics beyond the Standard Model (BSM). A broad class of proposed BSM models predicts new flavor-dependent interactions for leptons. Since neutrinos with different flavors experience these interactions differently, the oscillation patterns get modified, which can be seen in neutrino oscillation experiments. Over the years, various neutrino experiments have investigated the possibility of such BSM interactions and placed limits on their coupling strengths, including studies with atmospheric~\cite{Joshipura:2003jh, Khatun:2018lzs}, solar and reactor~\cite{Bandyopadhyay:2006uh, Gonzalez-Garcia:2006vic, Grifols:2003gy}, long-baseline~\cite{Chatterjee:2015gta, Singh:2023nek, Agarwalla:2024ylc, Heeck:2010pg, ESSnuSB:2025shd}, astrophysical neutrinos~\cite{Bustamante:2018mzu, Agarwalla:2023sng}, as well as with their combinations~\cite{Davoudiasl:2011sz, Farzan:2016wym, Wise:2018rnb, Heeck:2018nzc, Coloma:2020gfv, Dror:2020fbh, Alonso-Alvarez:2023tii}. Also, there are other complementary constraints on these BSM interactions from phenomena not involving neutrinos~\cite{Schlamminger:2007ht, Adelberger:2009zz, Salumbides:2013dua,Baryakhtar:2017ngi,KumarPoddar:2019ceq,KumarPoddar:2020kdz}. In this letter, we use weakly interacting neutrinos produced in the atmosphere of Earth to search for such lepton flavor-dependent BSM interactions. These atmospheric neutrinos span a wide energy range from a few MeV to TeV and traverse baselines from about 15 km to 12,750 km through Earth~\cite{Honda:2015fha}. The IceCube DeepCore detector at the South Pole, with its large volume and high-statistic atmospheric neutrino data, offers a unique opportunity to probe these BSM interactions.

\begin{figure}
	\centering
	\includegraphics[width=1\linewidth]{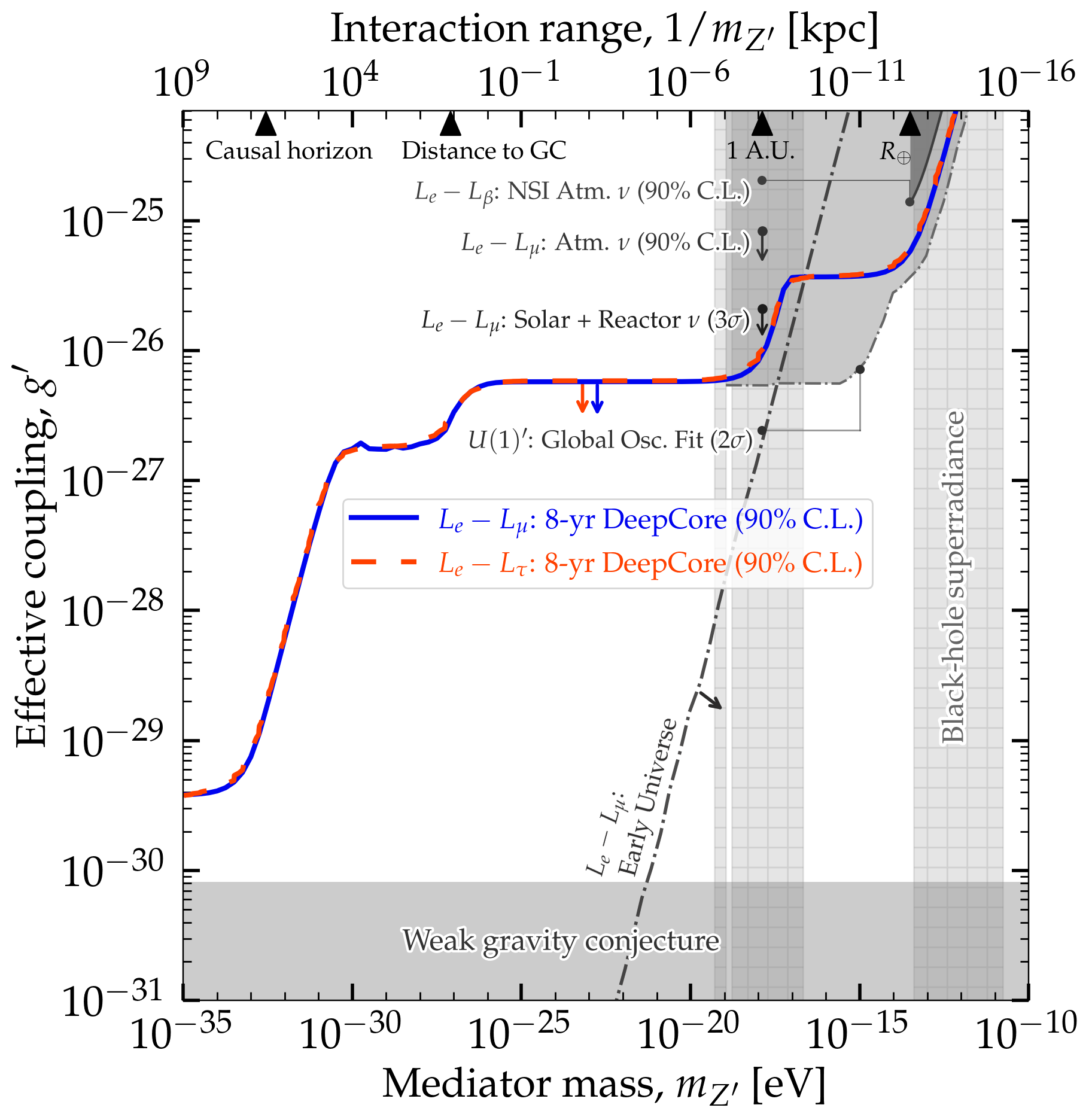}
	\caption{Constraints at the 90\% C.L. on LRI using an 8-year golden event sample of IceCube DeepCore~\cite{DVN_B4RITM_2025} in terms of the effective coupling strength $g'$ and the mass of the new gauge boson $Z'$ mediating LRI for ${L_e - L_\mu}$ (solid-blue curve) and ${L_e - L_\tau}$ (dashed-red curve) symmetries assuming normal mass ordering. The arrows indicate the allowed and the bands show the disallowed regions. See text for details regarding existing bounds shown here. These bounds on $g'$ and $m_Z'$ are derived from the limits on the LRI matter potentials (see Eq.~\ref{eq:LRI_potential_bounds}) following Refs.~\cite{Bustamante:2018mzu,Agarwalla:2023sng,Agarwalla:2024ylc}.}
	\label{fig:constraints}
\end{figure}

The SM contains accidental global $U(1)$ symmetries associated with baryon number $(B)$ and three lepton numbers ($L_e$, $L_\mu$, and $L_\tau$)~\cite{Langacker:2008yv}. Their certain linear combinations can be promoted to anomaly-free local symmetries, either within the SM particle content or with the addition of right-handed neutrinos. Among the simplest, well-motivated, and anomaly-free combinations are the lepton-number differences $L_\alpha - L_\beta$ (where $\alpha, \beta = e, \mu, \tau$)~\cite{He:1990pn, Foot:1990mn, He:1991qd, Foot:1994vd, Dutta:1994dx}.

In this work, we focus only on ${L_e - L_\mu}$ and ${L_e - L_\tau}$ symmetries sourced by electrons. Only one symmetry can be gauged at a time which introduces a new neutral gauge boson (\ie, $Z'_{e\mu}$ or  $Z'_{e\tau}$) that mediates vector-like flavor-dependent interactions between neutrinos and electrons. The phenomenological consequences of this new gauge boson depend strongly on its mass. A heavy gauge boson mediates short-range interactions that can be approximated as point-like neutrino interactions~\cite{Wolfenstein:1977ue, Valle:1987gv, Guzzo:1991hi, Langacker:2008yv, He:1991qd, Dutta:1994dx}, while an ultra-light one interacts with the abundant electrons in distant astrophysical environments, leading to a flavor-dependent long-range interaction (LRI) potential.

In this study, we specifically examine the ultra-light mediator scenario, where the interaction range is comparable to or exceeds the Sun-Earth distance. The strength of the induced LRI potential depends on the total number of electrons within the interaction range and the mass of the mediating gauge boson ($m_{Z'}$)~\cite{Joshipura:2003jh, Bandyopadhyay:2006uh, Gonzalez-Garcia:2006vic, Grifols:2003gy}. Such a long-range nature of these interactions allows even an extremely small coupling to accumulate a sizable amount of additional potential that can significantly alter neutrino oscillation probabilities. For example, taking the interaction range to be the Sun-Earth distance, $R_{SE} (\simeq 1.5 \times 10^{13}~\rm cm = 7.6 \times 10^{26}~GeV^{-1}$), the enormous number of electrons in the Sun $N_e (\sim 10^{57}$) generates a significant LRI potential at Earth~\cite{Joshipura:2003jh, Bandyopadhyay:2006uh} as,
\begin{equation}\label{eq:potential_coupling}
V_{e\mu/e\tau}(R_{SE}) = \alpha_{e\mu/e\tau}\frac{N_e}{R_{SE}} \approx 1.3 \times 10^{-11}~{\rm eV}
\left(\frac{\alpha_{e\mu/e\tau}}{10^{-50}}\right),
\end{equation}
where $\alpha_{e\mu/e\tau} \equiv {g'}^2_{e\mu/e\tau}/4\pi$ corresponds to the coupling strength of the new gauge boson. Figure~\ref{fig:constraints} presents the constraints on LRI potentials for ${L_e - L_\mu}$ (solid-blue curve) and ${L_e - L_\tau}$ (dashed-red curve) symmetries in terms of the corresponding effective coupling strength $g'$ and the mediator mass $m_{Z'}$ at 90\% confidence level (C.L.) using publicly available atmospheric neutrino data from IceCube DeepCore~\cite{DVN_B4RITM_2025}. We also show the comparison of our results with existing bounds from global oscillation fit~\cite{Coloma:2020gfv}, atmospheric neutrinos~\cite{Joshipura:2003jh}, solar and reactor neutrinos~\cite{Bandyopadhyay:2006uh}, and non-standard interaction searches~\cite{Super-Kamiokande:2011dam, Ohlsson:2012kf, Gonzalez-Garcia:2013usa}. Indirect limits~\cite{Wise:2018rnb} from black-hole superradiance~\cite{Baryakhtar:2017ngi}, the early Universe~\cite{Dror:2020fbh}, compact binaries~\cite{KumarPoddar:2019ceq}, and the weak gravity conjecture~\cite{Arkani-Hamed:2006emk} are also shown. These constraints from DeepCore are \emph{the most stringent direct experimental bound on such long-range interactions using neutrino data.} Now, we discuss the details of our analysis procedure.\\

\textbf{Effect of LRI on Neutrino Oscillations.---} 
The presence of LRI along with standard interactions (SI), referred to as the SI + LRI scenario, introduces an additional potential to the neutrino propagation Hamiltonian. The resulting Hamiltonian in the flavor basis is,
\begin{equation}
H_f=U\left[
\begin{tabular}{c c c} 0 & 0 & 0\\
0 & $\frac{\Delta m^2_{21}}{2E}$ & 0\\
0 & 0 & $\frac{\Delta m^2_{31}}{2E}$
\end{tabular}\right]U^\dag\, + \left[
\begin{tabular}{c c c} $V_{\rm CC}$ & 0 & 0\\ 0 & 0 & 0 \\ 0 & 0 & 0
\end{tabular} \right] + \left[
\begin{tabular}{c c c} $\zeta$ & 0 & 0 \\ 0 & $\xi$ & 0\\ 
0 & 0 & $\eta$ \\
\end{tabular} \right],\
\label{eq:modified_H}
\end{equation}
where the first term corresponds to vacuum oscillations governed by the Pontecorvo-Maki-Nakagawa-Sakata (PMNS) mixing matrix $U$~\cite{Maki:1962mu,Pontecorvo:1967fh} and the mass-squared splittings $\Delta m_{ij}^2$. The second term represents the contribution from the standard matter potential $V_{\rm CC}~ (= \pm \sqrt{2} G_F N_e$) arising from charged-current (CC) interaction of neutrinos with electrons in the Earth. Here, $G_F$ denotes the Fermi coupling constant and $N_e$ is the electron number density. The last term in Eq.~\ref{eq:modified_H} corresponds to the LRI contribution, which takes the form as, $(\zeta,\xi,\eta) = (V_{e\mu},-V_{e\mu},0)$ for $L_e-L_\mu$ symmetry and $(\zeta,\xi,\eta) = (V_{e\tau},0,-V_{e\tau})$ for $L_e-L_\tau$ symmetry.

\begin{figure*}[!t] 
	\centering
	\begin{minipage}[b]{1\textwidth}
		\includegraphics[width=\linewidth]{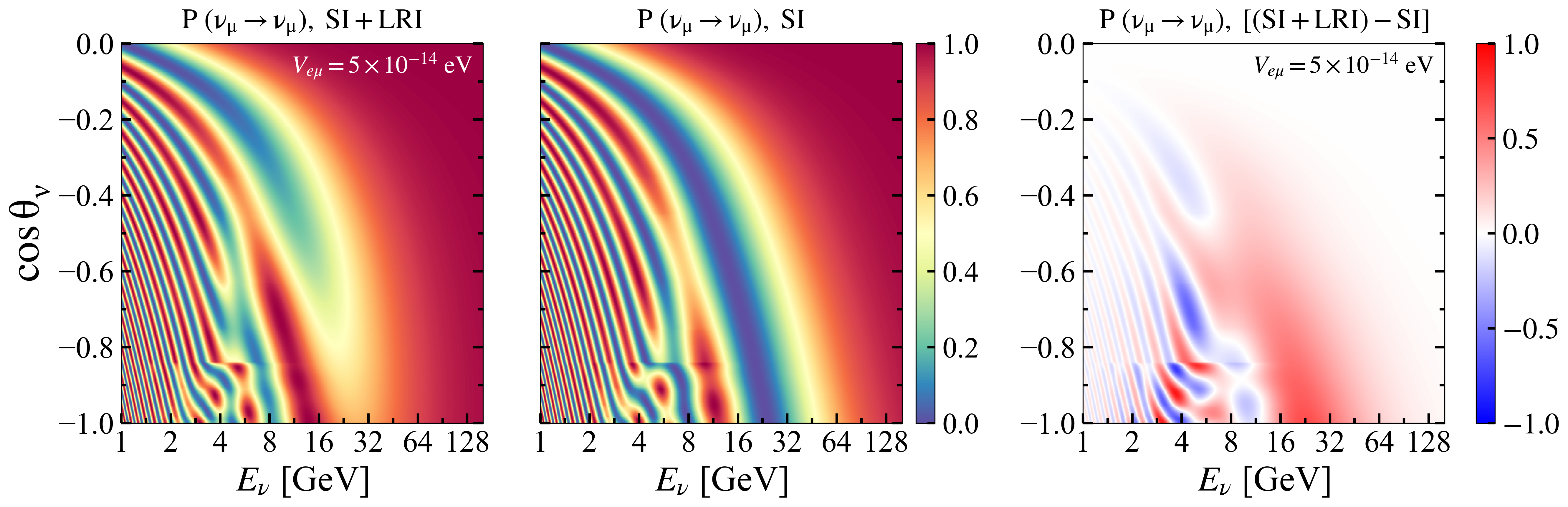}
	\end{minipage}
	\caption{$P(\nu_\mu \rightarrow \nu_\mu)$ oscillograms shown in $(E_{\nu},\,\cos\theta_{\nu})$ plane for SI and SI + LRI ($V_{e\mu} = 5 \times 10^{-14}\ \text{eV}$) scenarios in the middle and the left panels, respectively, assuming NO. The right panel illustrates the difference between the SI + LRI and SI scenarios. Here, we assume $\theta_{23} = 45.57^\circ$ and $\Delta m^2_{31} = 2.48 \times 10^{-3}~{\rm eV}^2$.
	}
	\label{fig:prob_V_emu}
\end{figure*}

In this work, we numerically calculate neutrino oscillation probabilities for normal mass ordering (NO) using a 12-layered density profile from the Preliminary Reference Earth Model (PREM)~\cite{Dziewonski:1981xy}. Figure~\ref{fig:prob_V_emu} shows the effect of the presence of LRI on the probability oscillogram for the $\nu_\mu \rightarrow \nu_\mu$ disappearance channel for a representative choice of $V_{e\mu} = 5 \times 10^{-14}\ \text{eV}$. The effect of LRI can be seen in terms of the disappearance of the oscillation valley (blue band) at longer baselines ($\cos\theta_{\nu} < -0.7$) around the energies of 16 GeV to 32 GeV in the left panel. The same effect appears as a prominent red region in the probability difference oscillogram in the right panel. This feature can be understood from the approximate dependence of effective mixing angle $\theta^m_{23}$ on $V_{e\mu}$~\cite{Chatterjee:2015gta} given by,
\begin{equation}
\tan 2\theta^m_{23} = \frac{\cos^{2}\theta_{13}
	- \alpha \cos^{2}\theta_{12}
	+ \alpha \sin^{2}\theta_{12} \sin^{2}\theta_{13}}
{W + \alpha \sin 2\theta_{12} \sin\theta_{13}},
\end{equation}
where $W \equiv 2EV_{e\mu}/\Delta m^2_{31}$ and $\alpha \equiv \Delta m^2_{21}/\Delta m^2_{31}$. In this expression, an increase in $V_{e\mu}$ suppresses $\theta^m_{23}$, which leads to the disappearance of the oscillation valley, as observed in Fig.~\ref{fig:prob_V_emu}. An analogous effect for the $L_e - L_\tau$ symmetry is discussed in the Supplemental Material. Now, we discuss the details of IceCube DeepCore.\\

\textbf{IceCube DeepCore Detector.---} IceCube is a cubic-kilometer neutrino observatory located at the geographic South Pole of the Earth~\cite{IceCube:2016zyt}. It consists of 5160 digital optical modules (DOMs)~\cite{IceCube:2010dpc}, deployed on 86 vertical strings embedded in the Antarctic ice at depths between 1450 m and 2450 m. The DOMs detect Cherenkov photons emitted by relativistic charged particles produced during neutrino interactions with ice. The $\nu_\mu$ CC interactions result in track-like events, whereas the $\nu_e$ CC, $\nu_\tau$ CC, and neutral current (NC) interactions of all flavors give rise to cascade-like events. A few $\nu_\tau$ CC interactions can also produce track-like signatures through muons originating from $\tau$ decay. The analysis is performed without differentiating between neutrinos and antineutrinos, as their event topologies are quite similar.

At the bottom-central region of IceCube lies DeepCore~\cite{IceCube:2011ucd}, a densely instrumented sub-array, where seven additional strings with higher quantum efficiency DOMs have been deployed. These features reduce the energy threshold of DeepCore to a few GeV, which is suitable for studying atmospheric neutrino oscillations. These studies include the precision measurements of the atmospheric mixing parameters ($\theta_{23}$ and $\Delta m^2_{32}$)~\cite{IceCube:2014flw, IceCube:2017lak, IceCubeCollaboration:2023wtb, IceCubeCollaboration:2024ssx}, tau neutrino appearance~\cite{IceCube:2019dqi}, and BSM searches~\cite{IceCube:2017ivd,IceCube:2020phf,IceCube:2020tka, IceCubeCollaboration:2024nle,IceCube:2024pky,IceCube:2024dlz, IceCube:2017zcu,IceCubeCollaboration:2021euf,IceCube:2025kve}.\\

\textbf{Events at IceCube DeepCore.---} In this study, we analyze the publicly available atmospheric neutrino sample of IceCube DeepCore collected during 2011-2019 with 7.5 years of livetime, having 21914 neutrino events in the reconstructed energy range of 6.3 - 158.5 GeV~\cite{IceCubeCollaboration:2023wtb,DVN_B4RITM_2025}. This data sample, referred to as ``\textit{golden event sample}'', excludes DOM hits where photons have experienced significant scatterings in ice. This is a significantly improved sample which employs updated optical module responses calibrated with in-situ data~\cite{IceCube:2020nwx}, an improved characterization of the glacial ice~\cite{Chirkin:2013tma}, updated simulations, an efficient event selection with enhanced background rejection, advanced event reconstruction techniques~\cite{IceCube:2022kff}, and a refined systematic treatment~\cite{IceCubeCollaboration:2023wtb}. 

To perform the analysis, we adopt a forward-folding approach using an extensive Monte Carlo (MC) simulation developed by the IceCube Collaboration. Each simulated events are reweighted using PISA framework~\cite{IceCube:2018ikn} to account for atmospheric neutrino flux, interaction cross sections, neutrino oscillations, and detector-response, ensuring that the sample accurately represents expected physical distributions in DeepCore. Various filters and reconstruction algorithms developed using this MC sample, have been applied to both simulated and observed events. To suppress the background from atmospheric muons and detector noise, multiple filters are applied, yielding a neutrino-dominated dataset. Event reconstruction algorithms based on maximum likelihood fits are used to reconstruct observables~\cite{IceCube:2014flw,Garza2014Measurement,AndriiThesis,IceCube:2022kff} like neutrino energy ($E_{\rm reco}$) and direction ($\cos\theta_{\rm reco}$). A Boosted Decision Tree (BDT)~\cite{Friedman:2001wbq} is trained to calculate a particle identification (PID) score representing the probability of an event being track-like. Further details regarding data acquisition, detector calibration, simulation, filters, and reconstruction have already been discussed in Ref.~\cite{IceCubeCollaboration:2023wtb}.

For this analysis, the data is binned in 10 logarithmic bins for $E_{\rm reco}$, covering the range 6.3 GeV - 158.5 GeV, and 10 linear bins for $\cos\theta_{\rm reco}$, spanning -1 to 0.1. The last energy bin is defined with twice the usual width to retain sufficient statistics. Subsequently, the events are categorized into two PID bins according to their event topology: mixed events ($0.25 < \rm PID < 0.55$) and track-like events ($0.55 < \rm PID < 1.0$). In order to achieve, high-purity $\nu_\mu$ CC sample, the cascade-like events with ${\rm PID} <0.25$ have been filtered out in this dataset.

\begin{figure}
	\centering
	\includegraphics[width=1.\linewidth]{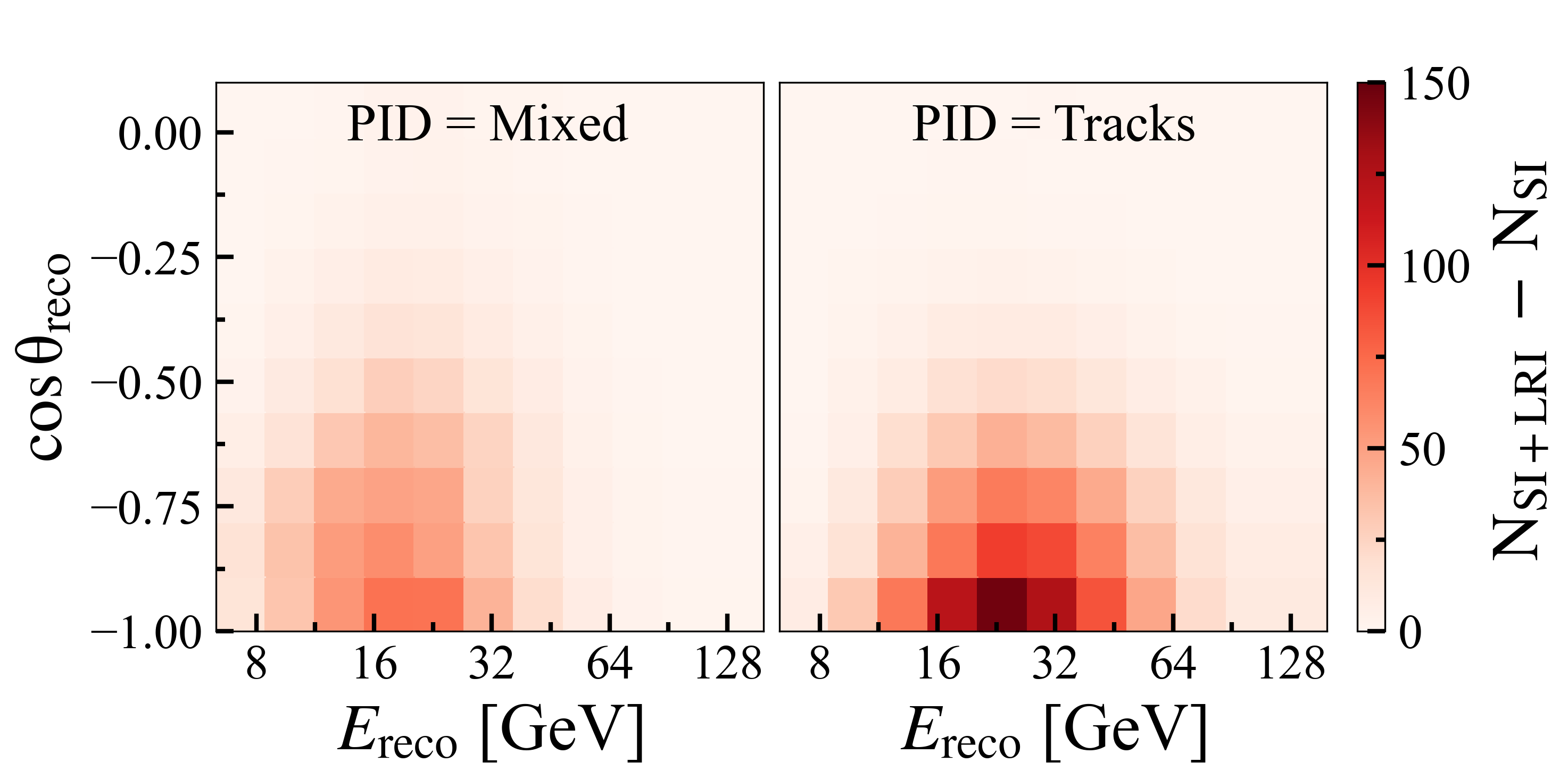}
	\caption{The difference of expected MC events at DeepCore for SI + LRI ($V_{e\mu} = 5 \times 10^{-14}$ eV) and SI scenarios. Here, we use the nominal values of the nuisance parameters as given in Table~\ref{tab:systematic_params}.
	}
	\label{fig:emu_event_diff}
\end{figure} 

\begin{figure*}[t]
	\centering
	\begin{minipage}[b]{0.49\textwidth}
		\includegraphics[width=\linewidth]{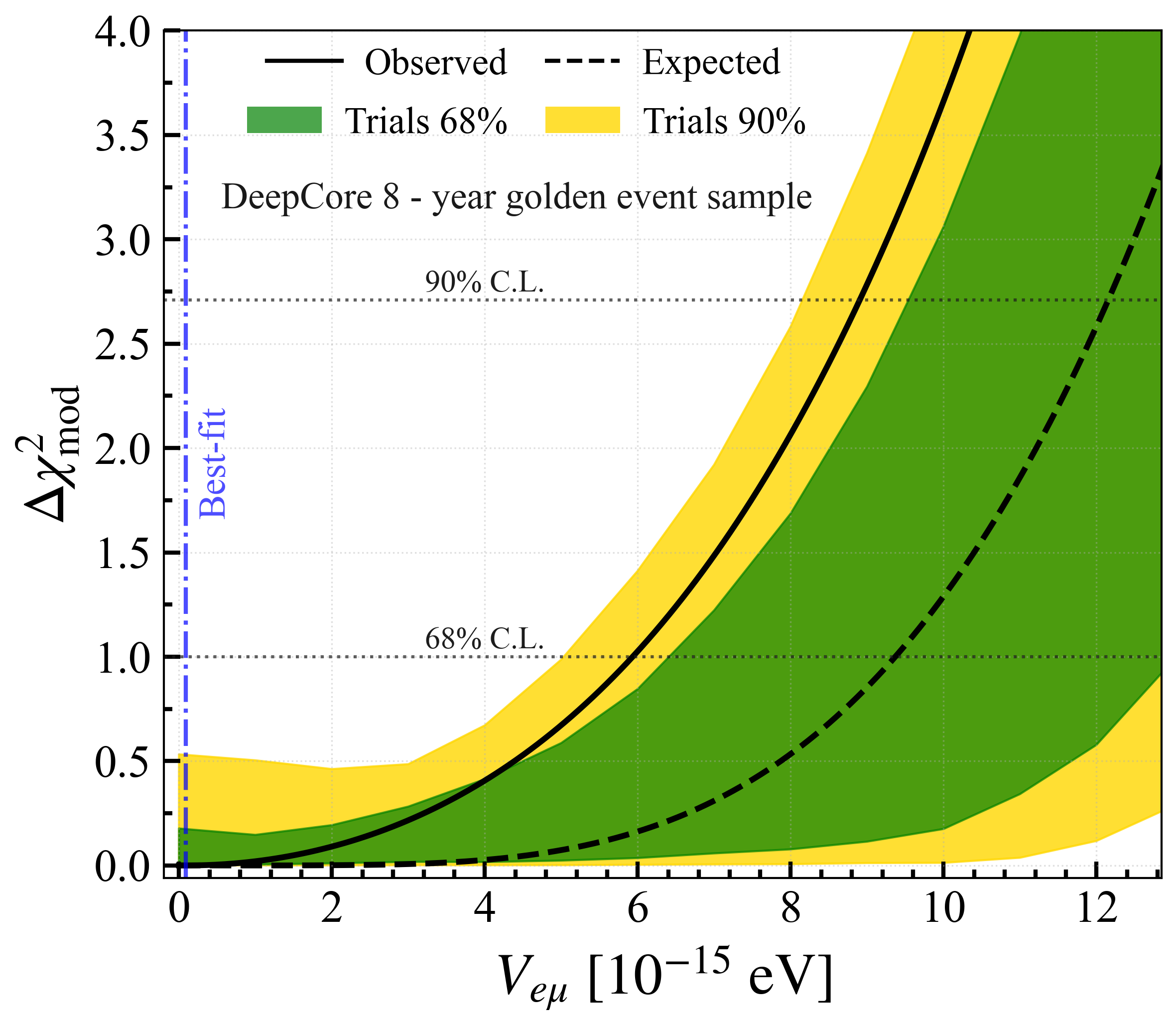}
	\end{minipage}
	\hfill
	\begin{minipage}[b]{0.49\textwidth}
		\includegraphics[width=\linewidth]{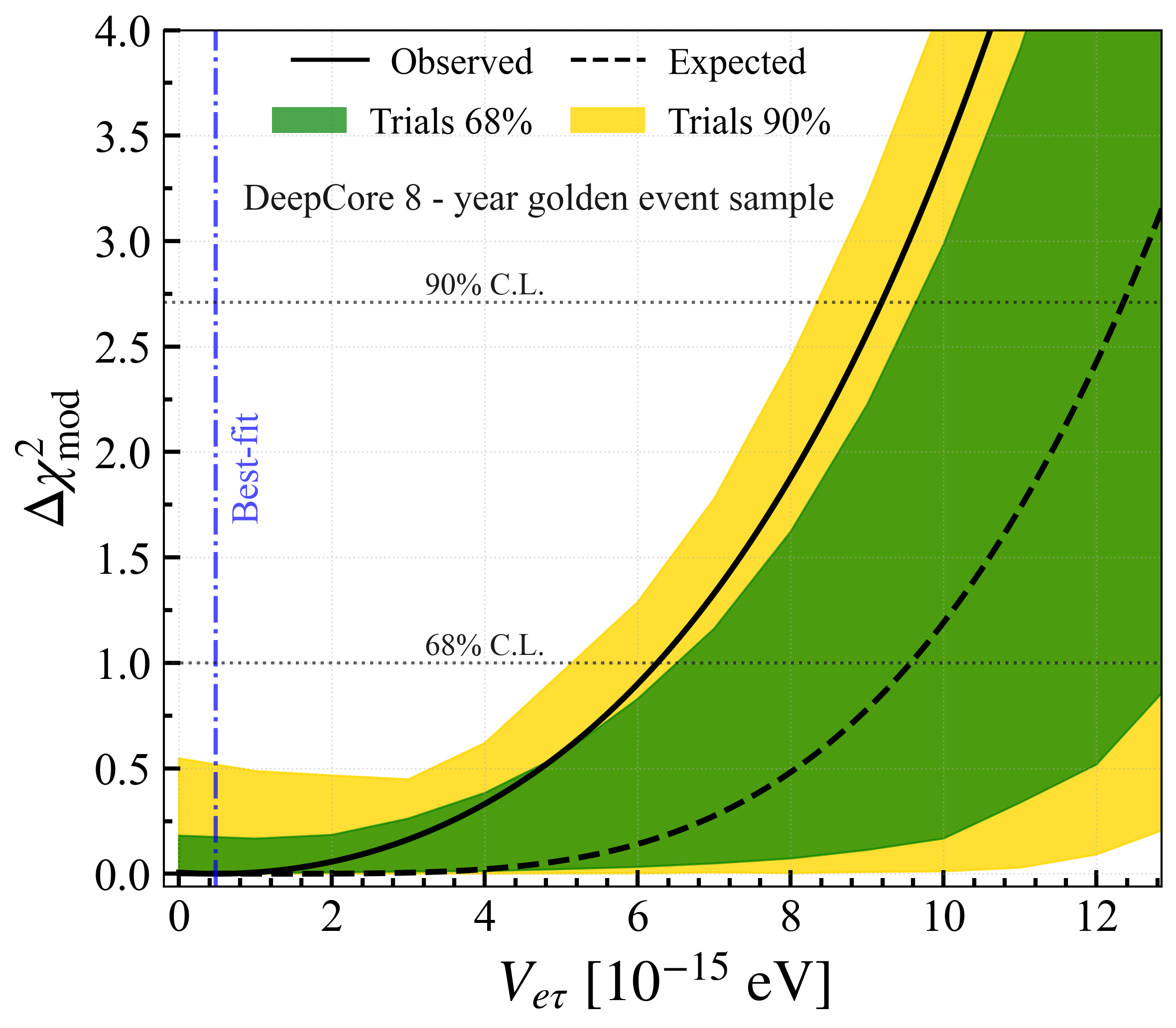}
	\end{minipage}
	\caption{Constraints on the long-range interaction potentials obtained using the 8-year golden event sample of IceCube DeepCore, assuming NO. The solid-black curves show the observed $\Delta \chi^2_{\rm mod}$ as a function of $V_{e\mu}$ (left) and $V_{e\tau}$ (right), while the dashed-black curves represent the corresponding expected sensitivities derived from the Asimov dataset. The vertical dash-dotted blue lines show the best-fit values of $V_{e\mu}$ (left) and $V_{e\tau}$ (right). The green (yellow) bands indicate the 68\% (90\%) containment regions from the distributions of $\Delta \chi^2_{\rm mod}$, which are obtained by fitting the statistically fluctuated 500 pseudo-trials.}
	\label{fig:results}
\end{figure*}

Figure~\ref{fig:emu_event_diff} shows the effect of $L_e - L_\mu$ symmetry on the nominal expected events. The left and right panels show the distributions of the difference of events for the SI + LRI and SI scenarios for mixed and track-like events, respectively. We can observe that the impact of LRI is prominent in the longer baseline and intermediate energy bins, which is consistent with the effects on oscillation probabilities as shown in Fig.~\ref{fig:prob_V_emu}.\\

\textbf{Numerical Analysis.---}  We perform this physics analysis by comparing the simulated event distribution with the observed data following the Frequentist method. In this analysis, we employ a modified $\chi^2$ as a test statistics following Ref.~\cite{IceCubeCollaboration:2023wtb}, which is given as,
\begin{equation} 
\chi^2_{\rm mod} = \sum_{i \in {\rm bins}} \frac{(N_i^{\rm exp} - N_i^{\rm obs})^2}{N_i^{\rm exp} + (\sigma_i^{\rm sim})^2} + \sum_{j \in {\rm syst}} \frac{(s_j - \hat{s}_j)^2}{\sigma^2_{s_j}}\, , \label{eq:mod_chi2}
\end{equation}
where $N_i^{\rm exp}$ and $N_i^{\rm obs}$ are the expected and observed event counts in the $i^{\rm th}$ analysis bin, respectively. The statistical uncertainty in $N_i^{\rm exp}$ due to the limited MC is given by $\sigma_i^{\rm sim}$. The last term corresponds to a Gaussian pull penalty for $j^{\rm th}$ systematic parameter with test value $s_j$ and a known uncertainty of $\sigma_{s_j}$ centered around its nominal value $\hat{s}_j$.

In our fit, we incorporate the systematic uncertainties associated with the atmospheric neutrino flux, oscillation parameters, interaction cross sections, detector response, overall neutrino normalization, and atmospheric muon normalization following the methodology described in Ref.~\cite{IceCubeCollaboration:2023wtb}. The complete list of all the free systematic parameters used in this analysis, along with their corresponding best-fit values, is provided in the Supplemental Material for both symmetries. As far as neutrino oscillation parameters are concerned, we fix $\theta_{12} = 33.41^\circ$, $\theta_{13} = 8.58^\circ$, and $\Delta m^2_{21} = 7.41 \times 10^{-5}~\text{eV}^2$, as reported by NuFit v5.2~\cite{Esteban:2020cvm}. Since the atmospheric data sample is not sensitive to the CP-violating phase $\delta_{\rm CP}$, it is fixed to 0. The remaining oscillation parameters $\theta_{23}$ and $\Delta m^2_{31}$ are minimized over the ranges $[38^\circ,\, 52^\circ]$ and $[2.0,\, 3.0] \times 10^{-3}~\text{eV}^2$, respectively, considering NO.\\

\textbf{Results.---} The left (right) panel of Fig.~\ref{fig:results} shows the constraints obtained from DeepCore on the LRI potential $V_{e\mu}$ ($V_{e\tau}$) corresponding to the $L_e-L_\mu$ ($L_e-L_\tau$) symmetry. We first fit the data with simulated MC events by varying the LRI parameter ($V_{\rm e\mu}$ or $V_{\rm e\tau}$, one-at-a-time) as well as all relevant nuisance parameters. The best-fit values of $V_{\rm e\mu}$ and $V_{\rm e\tau}$ obtained from the data fitting are
\begin{equation}
	V_{e\mu} = 8.9 \times 10^{-17}~{\rm eV}, 
	V_{e\tau} = 4.8 \times 10^{-16}~{\rm eV},
\end{equation}
which are consistent with the SI hypothesis with no LRI and indicated by the vertical dashed-dotted blue lines in the left and the right panels of Fig.~\ref{fig:results}, respectively. The best-fit values of all the systematic parameters obtained from the fitting are mentioned in the Table \ref{tab:systematic_params} of Supplemental Material.

Next, we perform a scan over the values of the LRI potential parameter $V_{\rm LRI}$ (\ie, $V_{\rm e\mu}$ or $V_{\rm e\tau}$, one-at-a-time), evaluating the corresponding $\Delta\chi^2_{\rm mod}$ defined as,
\begin{equation}
\Delta \chi^2_{\rm mod} = \chi^2_{\rm mod}(V_{\rm LRI}~{\rm fixed}) - \chi^2_{\rm mod}(V_{\rm LRI}~{\rm free})\,,
\end{equation}
where $\chi^2_{\rm mod}$ in both terms are minimized over the nuisance parameters. The solid-black curves in Fig.~\ref{fig:results} show the observed $\Delta \chi^2_{\rm mod}$, obtained from fitting the real data with the simulated data by profiling over $V_{\rm LRI}$. The 90\% C.L. upper bounds using these observed curves are: 
\begin{equation}\label{eq:LRI_potential_bounds}
V_{e\mu} < 8.9 \times 10^{-15}~{\rm eV}, ~ V_{e\tau} < 9.2 \times 10^{-15}~{\rm eV}.
\end{equation}
These upper bounds are used to obtain the constraints on the effective coupling strength and mediator mass as shown in Fig.~\ref{fig:constraints}. These results establish the most stringent constraints to date on flavor-dependent long-range neutrino interactions obtained from an atmospheric neutrino data analysis with a detailed systematic treatment. To evaluate the expected sensitivities shown by the dashed-black curves in Fig.~\ref{fig:results}, we generate an Asimov dataset at the best-fit values of $V_{\rm LRI}$ as well as nuisance parameters and calculate the expected $\Delta \chi^2_{\rm mod}$ by profiling over $V_{\rm LRI}$. The shaded bands represent the statistical uncertainties in the expected curve.\\

\textbf{Conclusion.---} In this work, we perform a comprehensive search for new flavor-dependent long-range interactions (LRI) of neutrinos arising from the gauged abelian lepton number symmetries, specifically, $L_e - L_\mu$ and $L_e - L_\tau$. To probe these LRI, we analyze for the first time in detail the 8-year high-purity $\nu_{\mu}$ CC atmospheric neutrino data collected by IceCube DeepCore. We find no evidence for LRI in this data sample and place the most stringent constraints on the coupling strength of these interactions for mediator masses lighter than $10^{-10}\,{\rm eV}$. The IceCube Upgrade~\cite{Ishihara:2019aao,IceCube:2025chb} currently under installation, ongoing KM3NeT/ORCA~\cite{KM3Net:2016zxf}, and the upcoming atmospheric neutrino detectors, such as DUNE~\cite{DUNE:2021cuw} and Hyper-K~\cite{Hyper-Kamiokande:2018ofw}, would further shed light on the properties of these ultra-light gauge bosons. \\

\begin{acknowledgments}
\textbf{Acknowledgments.}---  We acknowledge support from the Department of Atomic Energy (DAE), Govt. of India. G.G. acknowledges support from the Department of Science and Technology (DST), Govt. of India (Sanction Order No. DST/INSPIRE Fellowship/2021/IF210663). S.K.A., J.K., and A.K. receive support from the Swarnajayanti Fellowship (Sanction Order No. DST/SJF/PSA-05/2019-20) provided by the Department of Science and Technology (DST), Govt. of India, and the Research Grant (Sanction Order No. SB/SJF/2020-21/21) provided by the Anusandhan National Research Foundation (ANRF), Govt. of India, under the Swarnajayanti Fellowship. G.G. would like to thank A. Dighe and P. Swain for useful discussions. The numerical simulations are performed using the Dell PowerEdge R660 Server at the Institute of Physics,  Bhubaneswar, India.
\end{acknowledgments}

\bibliography{refs}

\providecommand{\noopsort}[1]{}\providecommand{\singleletter}[1]{#1}%
\begin{thebibliography}{82}%
\makeatletter
\providecommand \@ifxundefined [1]{%
 \@ifx{#1\undefined}
}%
\providecommand \@ifnum [1]{%
 \ifnum #1\expandafter \@firstoftwo
 \else \expandafter \@secondoftwo
 \fi
}%
\providecommand \@ifx [1]{%
 \ifx #1\expandafter \@firstoftwo
 \else \expandafter \@secondoftwo
 \fi
}%
\providecommand \natexlab [1]{#1}%
\providecommand \enquote  [1]{``#1''}%
\providecommand \bibnamefont  [1]{#1}%
\providecommand \bibfnamefont [1]{#1}%
\providecommand \citenamefont [1]{#1}%
\providecommand \href@noop [0]{\@secondoftwo}%
\providecommand \href [0]{\begingroup \@sanitize@url \@href}%
\providecommand \@href[1]{\@@startlink{#1}\@@href}%
\providecommand \@@href[1]{\endgroup#1\@@endlink}%
\providecommand \@sanitize@url [0]{\catcode `\\12\catcode `\$12\catcode
  `\&12\catcode `\#12\catcode `\^12\catcode `\_12\catcode `\%12\relax}%
\providecommand \@@startlink[1]{}%
\providecommand \@@endlink[0]{}%
\providecommand \url  [0]{\begingroup\@sanitize@url \@url }%
\providecommand \@url [1]{\endgroup\@href {#1}{\urlprefix }}%
\providecommand \urlprefix  [0]{URL }%
\providecommand \Eprint [0]{\href }%
\providecommand \doibase [0]{https://doi.org/}%
\providecommand \selectlanguage [0]{\@gobble}%
\providecommand \bibinfo  [0]{\@secondoftwo}%
\providecommand \bibfield  [0]{\@secondoftwo}%
\providecommand \translation [1]{[#1]}%
\providecommand \BibitemOpen [0]{}%
\providecommand \bibitemStop [0]{}%
\providecommand \bibitemNoStop [0]{.\EOS\space}%
\providecommand \EOS [0]{\spacefactor3000\relax}%
\providecommand \BibitemShut  [1]{\csname bibitem#1\endcsname}%
\let\auto@bib@innerbib\@empty
%</preamble>
\bibitem [{\citenamefont {Fukuda}\ \emph {et~al.}(1998)\citenamefont {Fukuda}
  \emph {et~al.}}]{Super-Kamiokande:1998kpq}%
  \BibitemOpen
  \bibfield  {author} {\bibinfo {author} {\bibfnamefont {Y.}~\bibnamefont
  {Fukuda}} \emph {et~al.} (\bibinfo {collaboration} {Super-Kamiokande}),\
  }\bibfield  {title} {\bibinfo {title} {{Evidence for oscillation of
  atmospheric neutrinos}},\ }\href
  {https://doi.org/10.1103/PhysRevLett.81.1562} {\bibfield  {journal} {\bibinfo
   {journal} {Phys. Rev. Lett.}\ }\textbf {\bibinfo {volume} {81}},\ \bibinfo
  {pages} {1562} (\bibinfo {year} {1998})},\ \Eprint
  {https://arxiv.org/abs/hep-ex/9807003} {arXiv:hep-ex/9807003} \BibitemShut
  {NoStop}%
\bibitem [{\citenamefont {Ahmad}\ \emph {et~al.}(2002)\citenamefont {Ahmad}
  \emph {et~al.}}]{SNO:2002tuh}%
  \BibitemOpen
  \bibfield  {author} {\bibinfo {author} {\bibfnamefont {Q.~R.}\ \bibnamefont
  {Ahmad}} \emph {et~al.} (\bibinfo {collaboration} {SNO}),\ }\bibfield
  {title} {\bibinfo {title} {{Direct evidence for neutrino flavor
  transformation from neutral current interactions in the Sudbury Neutrino
  Observatory}},\ }\href {https://doi.org/10.1103/PhysRevLett.89.011301}
  {\bibfield  {journal} {\bibinfo  {journal} {Phys. Rev. Lett.}\ }\textbf
  {\bibinfo {volume} {89}},\ \bibinfo {pages} {011301} (\bibinfo {year}
  {2002})},\ \Eprint {https://arxiv.org/abs/nucl-ex/0204008}
  {arXiv:nucl-ex/0204008} \BibitemShut {NoStop}%
\bibitem [{\citenamefont {Eguchi}\ \emph {et~al.}(2003)\citenamefont {Eguchi}
  \emph {et~al.}}]{KamLAND:2002uet}%
  \BibitemOpen
  \bibfield  {author} {\bibinfo {author} {\bibfnamefont {K.}~\bibnamefont
  {Eguchi}} \emph {et~al.} (\bibinfo {collaboration} {KamLAND}),\ }\bibfield
  {title} {\bibinfo {title} {{First results from KamLAND: Evidence for reactor
  anti-neutrino disappearance}},\ }\href
  {https://doi.org/10.1103/PhysRevLett.90.021802} {\bibfield  {journal}
  {\bibinfo  {journal} {Phys. Rev. Lett.}\ }\textbf {\bibinfo {volume} {90}},\
  \bibinfo {pages} {021802} (\bibinfo {year} {2003})},\ \Eprint
  {https://arxiv.org/abs/hep-ex/0212021} {arXiv:hep-ex/0212021} \BibitemShut
  {NoStop}%
\bibitem [{\citenamefont {An}\ \emph {et~al.}(2012)\citenamefont {An} \emph
  {et~al.}}]{DayaBay:2012fng}%
  \BibitemOpen
  \bibfield  {author} {\bibinfo {author} {\bibfnamefont {F.~P.}\ \bibnamefont
  {An}} \emph {et~al.} (\bibinfo {collaboration} {Daya Bay}),\ }\bibfield
  {title} {\bibinfo {title} {{Observation of electron-antineutrino
  disappearance at Daya Bay}},\ }\href
  {https://doi.org/10.1103/PhysRevLett.108.171803} {\bibfield  {journal}
  {\bibinfo  {journal} {Phys. Rev. Lett.}\ }\textbf {\bibinfo {volume} {108}},\
  \bibinfo {pages} {171803} (\bibinfo {year} {2012})},\ \Eprint
  {https://arxiv.org/abs/1203.1669} {arXiv:1203.1669 [hep-ex]} \BibitemShut
  {NoStop}%
\bibitem [{\citenamefont {Aartsen}\ \emph {et~al.}(2015)\citenamefont {Aartsen}
  \emph {et~al.}}]{IceCube:2014flw}%
  \BibitemOpen
  \bibfield  {author} {\bibinfo {author} {\bibfnamefont {M.~G.}\ \bibnamefont
  {Aartsen}} \emph {et~al.} (\bibinfo {collaboration} {IceCube}),\ }\bibfield
  {title} {\bibinfo {title} {{Determining neutrino oscillation parameters from
  atmospheric muon neutrino disappearance with three years of IceCube DeepCore
  data}},\ }\href {https://doi.org/10.1103/PhysRevD.91.072004} {\bibfield
  {journal} {\bibinfo  {journal} {Phys. Rev. D}\ }\textbf {\bibinfo {volume}
  {91}},\ \bibinfo {pages} {072004} (\bibinfo {year} {2015})},\ \Eprint
  {https://arxiv.org/abs/1410.7227} {arXiv:1410.7227 [hep-ex]} \BibitemShut
  {NoStop}%
\bibitem [{\citenamefont {Joshipura}\ and\ \citenamefont
  {Mohanty}(2004)}]{Joshipura:2003jh}%
  \BibitemOpen
  \bibfield  {author} {\bibinfo {author} {\bibfnamefont {A.~S.}\ \bibnamefont
  {Joshipura}}\ and\ \bibinfo {author} {\bibfnamefont {S.}~\bibnamefont
  {Mohanty}},\ }\bibfield  {title} {\bibinfo {title} {{Constraints on flavor
  dependent long range forces from atmospheric neutrino observations at
  super-Kamiokande}},\ }\href {https://doi.org/10.1016/j.physletb.2004.01.057}
  {\bibfield  {journal} {\bibinfo  {journal} {Phys. Lett. B}\ }\textbf
  {\bibinfo {volume} {584}},\ \bibinfo {pages} {103} (\bibinfo {year}
  {2004})},\ \Eprint {https://arxiv.org/abs/hep-ph/0310210}
  {arXiv:hep-ph/0310210} \BibitemShut {NoStop}%
\bibitem [{\citenamefont {Khatun}\ \emph {et~al.}(2018)\citenamefont {Khatun},
  \citenamefont {Thakore},\ and\ \citenamefont {Agarwalla}}]{Khatun:2018lzs}%
  \BibitemOpen
  \bibfield  {author} {\bibinfo {author} {\bibfnamefont {A.}~\bibnamefont
  {Khatun}}, \bibinfo {author} {\bibfnamefont {T.}~\bibnamefont {Thakore}},\
  and\ \bibinfo {author} {\bibfnamefont {S.~K.}\ \bibnamefont {Agarwalla}},\
  }\bibfield  {title} {\bibinfo {title} {{Can INO be Sensitive to
  Flavor-Dependent Long-Range Forces?}},\ }\href
  {https://doi.org/10.1007/JHEP04(2018)023} {\bibfield  {journal} {\bibinfo
  {journal} {JHEP}\ }\textbf {\bibinfo {volume} {04}},\ \bibinfo {pages}
  {023}},\ \Eprint {https://arxiv.org/abs/1801.00949} {arXiv:1801.00949
  [hep-ph]} \BibitemShut {NoStop}%
\bibitem [{\citenamefont {Bandyopadhyay}\ \emph {et~al.}(2007)\citenamefont
  {Bandyopadhyay}, \citenamefont {Dighe},\ and\ \citenamefont
  {Joshipura}}]{Bandyopadhyay:2006uh}%
  \BibitemOpen
  \bibfield  {author} {\bibinfo {author} {\bibfnamefont {A.}~\bibnamefont
  {Bandyopadhyay}}, \bibinfo {author} {\bibfnamefont {A.}~\bibnamefont
  {Dighe}},\ and\ \bibinfo {author} {\bibfnamefont {A.~S.}\ \bibnamefont
  {Joshipura}},\ }\bibfield  {title} {\bibinfo {title} {{Constraints on
  flavor-dependent long range forces from solar neutrinos and KamLAND}},\
  }\href {https://doi.org/10.1103/PhysRevD.75.093005} {\bibfield  {journal}
  {\bibinfo  {journal} {Phys. Rev. D}\ }\textbf {\bibinfo {volume} {75}},\
  \bibinfo {pages} {093005} (\bibinfo {year} {2007})},\ \Eprint
  {https://arxiv.org/abs/hep-ph/0610263} {arXiv:hep-ph/0610263} \BibitemShut
  {NoStop}%
\bibitem [{\citenamefont {Gonzalez-Garcia}\ \emph {et~al.}(2007)\citenamefont
  {Gonzalez-Garcia}, \citenamefont {de~Holanda}, \citenamefont {Masso},\ and\
  \citenamefont {Zukanovich~Funchal}}]{Gonzalez-Garcia:2006vic}%
  \BibitemOpen
  \bibfield  {author} {\bibinfo {author} {\bibfnamefont {M.~C.}\ \bibnamefont
  {Gonzalez-Garcia}}, \bibinfo {author} {\bibfnamefont {P.~C.}\ \bibnamefont
  {de~Holanda}}, \bibinfo {author} {\bibfnamefont {E.}~\bibnamefont {Masso}},\
  and\ \bibinfo {author} {\bibfnamefont {R.}~\bibnamefont
  {Zukanovich~Funchal}},\ }\bibfield  {title} {\bibinfo {title} {{Probing
  long-range leptonic forces with solar and reactor neutrinos}},\ }\href
  {https://doi.org/10.1088/1475-7516/2007/01/005} {\bibfield  {journal}
  {\bibinfo  {journal} {JCAP}\ }\textbf {\bibinfo {volume} {01}},\ \bibinfo
  {pages} {005}},\ \Eprint {https://arxiv.org/abs/hep-ph/0609094}
  {arXiv:hep-ph/0609094} \BibitemShut {NoStop}%
\bibitem [{\citenamefont {Grifols}\ and\ \citenamefont
  {Masso}(2004)}]{Grifols:2003gy}%
  \BibitemOpen
  \bibfield  {author} {\bibinfo {author} {\bibfnamefont {J.~A.}\ \bibnamefont
  {Grifols}}\ and\ \bibinfo {author} {\bibfnamefont {E.}~\bibnamefont
  {Masso}},\ }\bibfield  {title} {\bibinfo {title} {{Neutrino oscillations in
  the sun probe long range leptonic forces}},\ }\href
  {https://doi.org/10.1016/j.physletb.2003.10.078} {\bibfield  {journal}
  {\bibinfo  {journal} {Phys. Lett. B}\ }\textbf {\bibinfo {volume} {579}},\
  \bibinfo {pages} {123} (\bibinfo {year} {2004})},\ \Eprint
  {https://arxiv.org/abs/hep-ph/0311141} {arXiv:hep-ph/0311141} \BibitemShut
  {NoStop}%
\bibitem [{\citenamefont {Chatterjee}\ \emph {et~al.}(2015)\citenamefont
  {Chatterjee}, \citenamefont {Dasgupta},\ and\ \citenamefont
  {Agarwalla}}]{Chatterjee:2015gta}%
  \BibitemOpen
  \bibfield  {author} {\bibinfo {author} {\bibfnamefont {S.~S.}\ \bibnamefont
  {Chatterjee}}, \bibinfo {author} {\bibfnamefont {A.}~\bibnamefont
  {Dasgupta}},\ and\ \bibinfo {author} {\bibfnamefont {S.~K.}\ \bibnamefont
  {Agarwalla}},\ }\bibfield  {title} {\bibinfo {title} {{Exploring
  Flavor-Dependent Long-Range Forces in Long-Baseline Neutrino Oscillation
  Experiments}},\ }\href {https://doi.org/10.1007/JHEP12(2015)167} {\bibfield
  {journal} {\bibinfo  {journal} {JHEP}\ }\textbf {\bibinfo {volume} {12}},\
  \bibinfo {pages} {167}},\ \Eprint {https://arxiv.org/abs/1509.03517}
  {arXiv:1509.03517 [hep-ph]} \BibitemShut {NoStop}%
\bibitem [{\citenamefont {Singh}\ \emph {et~al.}(2023)\citenamefont {Singh},
  \citenamefont {Bustamante},\ and\ \citenamefont {Agarwalla}}]{Singh:2023nek}%
  \BibitemOpen
  \bibfield  {author} {\bibinfo {author} {\bibfnamefont {M.}~\bibnamefont
  {Singh}}, \bibinfo {author} {\bibfnamefont {M.}~\bibnamefont {Bustamante}},\
  and\ \bibinfo {author} {\bibfnamefont {S.~K.}\ \bibnamefont {Agarwalla}},\
  }\bibfield  {title} {\bibinfo {title} {{Flavor-dependent long-range neutrino
  interactions in DUNE {\&} T2HK: alone they constrain, together they
  discover}},\ }\href {https://doi.org/10.1007/JHEP08(2023)101} {\bibfield
  {journal} {\bibinfo  {journal} {JHEP}\ }\textbf {\bibinfo {volume} {08}},\
  \bibinfo {pages} {101}},\ \Eprint {https://arxiv.org/abs/2305.05184}
  {arXiv:2305.05184 [hep-ph]} \BibitemShut {NoStop}%
\bibitem [{\citenamefont {Agarwalla}\ \emph {et~al.}(2024)\citenamefont
  {Agarwalla}, \citenamefont {Bustamante}, \citenamefont {Singh},\ and\
  \citenamefont {Swain}}]{Agarwalla:2024ylc}%
  \BibitemOpen
  \bibfield  {author} {\bibinfo {author} {\bibfnamefont {S.~K.}\ \bibnamefont
  {Agarwalla}}, \bibinfo {author} {\bibfnamefont {M.}~\bibnamefont
  {Bustamante}}, \bibinfo {author} {\bibfnamefont {M.}~\bibnamefont {Singh}},\
  and\ \bibinfo {author} {\bibfnamefont {P.}~\bibnamefont {Swain}},\ }\bibfield
   {title} {\bibinfo {title} {{A plethora of long-range neutrino interactions
  probed by DUNE and T2HK}},\ }\href {https://doi.org/10.1007/JHEP09(2024)055}
  {\bibfield  {journal} {\bibinfo  {journal} {JHEP}\ }\textbf {\bibinfo
  {volume} {09}},\ \bibinfo {pages} {055}},\ \Eprint
  {https://arxiv.org/abs/2404.02775} {arXiv:2404.02775 [hep-ph]} \BibitemShut
  {NoStop}%
\bibitem [{\citenamefont {Heeck}\ and\ \citenamefont
  {Rodejohann}(2011)}]{Heeck:2010pg}%
  \BibitemOpen
  \bibfield  {author} {\bibinfo {author} {\bibfnamefont {J.}~\bibnamefont
  {Heeck}}\ and\ \bibinfo {author} {\bibfnamefont {W.}~\bibnamefont
  {Rodejohann}},\ }\bibfield  {title} {\bibinfo {title} {{Gauged $L_\mu -
  L_\tau$ and different Muon Neutrino and Anti-Neutrino Oscillations: MINOS and
  beyond}},\ }\href {https://doi.org/10.1088/0954-3899/38/8/085005} {\bibfield
  {journal} {\bibinfo  {journal} {J. Phys. G}\ }\textbf {\bibinfo {volume}
  {38}},\ \bibinfo {pages} {085005} (\bibinfo {year} {2011})},\ \Eprint
  {https://arxiv.org/abs/1007.2655} {arXiv:1007.2655 [hep-ph]} \BibitemShut
  {NoStop}%
\bibitem [{\citenamefont {Aguilar}\ \emph {et~al.}(2025)\citenamefont {Aguilar}
  \emph {et~al.}}]{ESSnuSB:2025shd}%
  \BibitemOpen
  \bibfield  {author} {\bibinfo {author} {\bibfnamefont {J.}~\bibnamefont
  {Aguilar}} \emph {et~al.} (\bibinfo {collaboration} {ESSnuSB}),\ }\bibfield
  {title} {\bibinfo {title} {{Probing long-range forces in neutrino
  oscillations at the ESSnuSB experiment}},\ }\href
  {https://doi.org/10.1007/JHEP07(2025)186} {\bibfield  {journal} {\bibinfo
  {journal} {JHEP}\ }\textbf {\bibinfo {volume} {07}},\ \bibinfo {pages}
  {186}},\ \Eprint {https://arxiv.org/abs/2504.10480} {arXiv:2504.10480
  [hep-ph]} \BibitemShut {NoStop}%
\bibitem [{\citenamefont {Bustamante}\ and\ \citenamefont
  {Agarwalla}(2019)}]{Bustamante:2018mzu}%
  \BibitemOpen
  \bibfield  {author} {\bibinfo {author} {\bibfnamefont {M.}~\bibnamefont
  {Bustamante}}\ and\ \bibinfo {author} {\bibfnamefont {S.~K.}\ \bibnamefont
  {Agarwalla}},\ }\bibfield  {title} {\bibinfo {title} {{Universe's Worth of
  Electrons to Probe Long-Range Interactions of High-Energy Astrophysical
  Neutrinos}},\ }\href {https://doi.org/10.1103/PhysRevLett.122.061103}
  {\bibfield  {journal} {\bibinfo  {journal} {Phys. Rev. Lett.}\ }\textbf
  {\bibinfo {volume} {122}},\ \bibinfo {pages} {061103} (\bibinfo {year}
  {2019})},\ \Eprint {https://arxiv.org/abs/1808.02042} {arXiv:1808.02042
  [astro-ph.HE]} \BibitemShut {NoStop}%
\bibitem [{\citenamefont {Agarwalla}\ \emph {et~al.}(2023)\citenamefont
  {Agarwalla}, \citenamefont {Bustamante}, \citenamefont {Das},\ and\
  \citenamefont {Narang}}]{Agarwalla:2023sng}%
  \BibitemOpen
  \bibfield  {author} {\bibinfo {author} {\bibfnamefont {S.~K.}\ \bibnamefont
  {Agarwalla}}, \bibinfo {author} {\bibfnamefont {M.}~\bibnamefont
  {Bustamante}}, \bibinfo {author} {\bibfnamefont {S.}~\bibnamefont {Das}},\
  and\ \bibinfo {author} {\bibfnamefont {A.}~\bibnamefont {Narang}},\
  }\bibfield  {title} {\bibinfo {title} {{Present and future constraints on
  flavor-dependent long-range interactions of high-energy astrophysical
  neutrinos}},\ }\href {https://doi.org/10.1007/JHEP08(2023)113} {\bibfield
  {journal} {\bibinfo  {journal} {JHEP}\ }\textbf {\bibinfo {volume} {08}},\
  \bibinfo {pages} {113}},\ \Eprint {https://arxiv.org/abs/2305.03675}
  {arXiv:2305.03675 [hep-ph]} \BibitemShut {NoStop}%
\bibitem [{\citenamefont {Davoudiasl}\ \emph {et~al.}(2011)\citenamefont
  {Davoudiasl}, \citenamefont {Lee},\ and\ \citenamefont
  {Marciano}}]{Davoudiasl:2011sz}%
  \BibitemOpen
  \bibfield  {author} {\bibinfo {author} {\bibfnamefont {H.}~\bibnamefont
  {Davoudiasl}}, \bibinfo {author} {\bibfnamefont {H.-S.}\ \bibnamefont
  {Lee}},\ and\ \bibinfo {author} {\bibfnamefont {W.~J.}\ \bibnamefont
  {Marciano}},\ }\bibfield  {title} {\bibinfo {title} {{Long-Range Lepton
  Flavor Interactions and Neutrino Oscillations}},\ }\href
  {https://doi.org/10.1103/PhysRevD.84.013009} {\bibfield  {journal} {\bibinfo
  {journal} {Phys. Rev. D}\ }\textbf {\bibinfo {volume} {84}},\ \bibinfo
  {pages} {013009} (\bibinfo {year} {2011})},\ \Eprint
  {https://arxiv.org/abs/1102.5352} {arXiv:1102.5352 [hep-ph]} \BibitemShut
  {NoStop}%
\bibitem [{\citenamefont {Farzan}\ and\ \citenamefont
  {Heeck}(2016)}]{Farzan:2016wym}%
  \BibitemOpen
  \bibfield  {author} {\bibinfo {author} {\bibfnamefont {Y.}~\bibnamefont
  {Farzan}}\ and\ \bibinfo {author} {\bibfnamefont {J.}~\bibnamefont {Heeck}},\
  }\bibfield  {title} {\bibinfo {title} {{Neutrinophilic nonstandard
  interactions}},\ }\href {https://doi.org/10.1103/PhysRevD.94.053010}
  {\bibfield  {journal} {\bibinfo  {journal} {Phys. Rev. D}\ }\textbf {\bibinfo
  {volume} {94}},\ \bibinfo {pages} {053010} (\bibinfo {year} {2016})},\
  \Eprint {https://arxiv.org/abs/1607.07616} {arXiv:1607.07616 [hep-ph]}
  \BibitemShut {NoStop}%
\bibitem [{\citenamefont {Wise}\ and\ \citenamefont
  {Zhang}(2018)}]{Wise:2018rnb}%
  \BibitemOpen
  \bibfield  {author} {\bibinfo {author} {\bibfnamefont {M.~B.}\ \bibnamefont
  {Wise}}\ and\ \bibinfo {author} {\bibfnamefont {Y.}~\bibnamefont {Zhang}},\
  }\bibfield  {title} {\bibinfo {title} {{Lepton Flavorful Fifth Force and
  Depth-dependent Neutrino Matter Interactions}},\ }\href
  {https://doi.org/10.1007/JHEP06(2018)053} {\bibfield  {journal} {\bibinfo
  {journal} {JHEP}\ }\textbf {\bibinfo {volume} {06}},\ \bibinfo {pages}
  {053}},\ \Eprint {https://arxiv.org/abs/1803.00591} {arXiv:1803.00591
  [hep-ph]} \BibitemShut {NoStop}%
\bibitem [{\citenamefont {Heeck}\ \emph {et~al.}(2019)\citenamefont {Heeck},
  \citenamefont {Lindner}, \citenamefont {Rodejohann},\ and\ \citenamefont
  {Vogl}}]{Heeck:2018nzc}%
  \BibitemOpen
  \bibfield  {author} {\bibinfo {author} {\bibfnamefont {J.}~\bibnamefont
  {Heeck}}, \bibinfo {author} {\bibfnamefont {M.}~\bibnamefont {Lindner}},
  \bibinfo {author} {\bibfnamefont {W.}~\bibnamefont {Rodejohann}},\ and\
  \bibinfo {author} {\bibfnamefont {S.}~\bibnamefont {Vogl}},\ }\bibfield
  {title} {\bibinfo {title} {{Non-Standard Neutrino Interactions and Neutral
  Gauge Bosons}},\ }\href {https://doi.org/10.21468/SciPostPhys.6.3.038}
  {\bibfield  {journal} {\bibinfo  {journal} {SciPost Phys.}\ }\textbf
  {\bibinfo {volume} {6}},\ \bibinfo {pages} {038} (\bibinfo {year} {2019})},\
  \Eprint {https://arxiv.org/abs/1812.04067} {arXiv:1812.04067 [hep-ph]}
  \BibitemShut {NoStop}%
\bibitem [{\citenamefont {Coloma}\ \emph {et~al.}(2021)\citenamefont {Coloma},
  \citenamefont {Gonzalez-Garcia},\ and\ \citenamefont
  {Maltoni}}]{Coloma:2020gfv}%
  \BibitemOpen
  \bibfield  {author} {\bibinfo {author} {\bibfnamefont {P.}~\bibnamefont
  {Coloma}}, \bibinfo {author} {\bibfnamefont {M.~C.}\ \bibnamefont
  {Gonzalez-Garcia}},\ and\ \bibinfo {author} {\bibfnamefont {M.}~\bibnamefont
  {Maltoni}},\ }\bibfield  {title} {\bibinfo {title} {{Neutrino oscillation
  constraints on U(1)' models: from non-standard interactions to long-range
  forces}},\ }\href {https://doi.org/10.1007/JHEP01(2021)114} {\bibfield
  {journal} {\bibinfo  {journal} {JHEP}\ }\textbf {\bibinfo {volume} {01}},\
  \bibinfo {pages} {114}},\ \bibinfo {note} {[Erratum: JHEP 11, 115 (2022)]},\
  \Eprint {https://arxiv.org/abs/2009.14220} {arXiv:2009.14220 [hep-ph]}
  \BibitemShut {NoStop}%
\bibitem [{\citenamefont {Dror}(2020)}]{Dror:2020fbh}%
  \BibitemOpen
  \bibfield  {author} {\bibinfo {author} {\bibfnamefont {J.~A.}\ \bibnamefont
  {Dror}},\ }\bibfield  {title} {\bibinfo {title} {{Discovering leptonic forces
  using nonconserved currents}},\ }\href
  {https://doi.org/10.1103/PhysRevD.101.095013} {\bibfield  {journal} {\bibinfo
   {journal} {Phys. Rev. D}\ }\textbf {\bibinfo {volume} {101}},\ \bibinfo
  {pages} {095013} (\bibinfo {year} {2020})},\ \Eprint
  {https://arxiv.org/abs/2004.04750} {arXiv:2004.04750 [hep-ph]} \BibitemShut
  {NoStop}%
\bibitem [{\citenamefont {Alonso-{\'A}lvarez}\ \emph
  {et~al.}(2023)\citenamefont {Alonso-{\'A}lvarez}, \citenamefont {Bleau},\
  and\ \citenamefont {Cline}}]{Alonso-Alvarez:2023tii}%
  \BibitemOpen
  \bibfield  {author} {\bibinfo {author} {\bibfnamefont {G.}~\bibnamefont
  {Alonso-{\'A}lvarez}}, \bibinfo {author} {\bibfnamefont {K.}~\bibnamefont
  {Bleau}},\ and\ \bibinfo {author} {\bibfnamefont {J.~M.}\ \bibnamefont
  {Cline}},\ }\bibfield  {title} {\bibinfo {title} {{Distortion of neutrino
  oscillations by dark photon dark matter}},\ }\href
  {https://doi.org/10.1103/PhysRevD.107.055045} {\bibfield  {journal} {\bibinfo
   {journal} {Phys. Rev. D}\ }\textbf {\bibinfo {volume} {107}},\ \bibinfo
  {pages} {055045} (\bibinfo {year} {2023})},\ \Eprint
  {https://arxiv.org/abs/2301.04152} {arXiv:2301.04152 [hep-ph]} \BibitemShut
  {NoStop}%
\bibitem [{\citenamefont {Schlamminger}\ \emph {et~al.}(2008)\citenamefont
  {Schlamminger}, \citenamefont {Choi}, \citenamefont {Wagner}, \citenamefont
  {Gundlach},\ and\ \citenamefont {Adelberger}}]{Schlamminger:2007ht}%
  \BibitemOpen
  \bibfield  {author} {\bibinfo {author} {\bibfnamefont {S.}~\bibnamefont
  {Schlamminger}}, \bibinfo {author} {\bibfnamefont {K.~Y.}\ \bibnamefont
  {Choi}}, \bibinfo {author} {\bibfnamefont {T.~A.}\ \bibnamefont {Wagner}},
  \bibinfo {author} {\bibfnamefont {J.~H.}\ \bibnamefont {Gundlach}},\ and\
  \bibinfo {author} {\bibfnamefont {E.~G.}\ \bibnamefont {Adelberger}},\
  }\bibfield  {title} {\bibinfo {title} {{Test of the equivalence principle
  using a rotating torsion balance}},\ }\href
  {https://doi.org/10.1103/PhysRevLett.100.041101} {\bibfield  {journal}
  {\bibinfo  {journal} {Phys. Rev. Lett.}\ }\textbf {\bibinfo {volume} {100}},\
  \bibinfo {pages} {041101} (\bibinfo {year} {2008})},\ \Eprint
  {https://arxiv.org/abs/0712.0607} {arXiv:0712.0607 [gr-qc]} \BibitemShut
  {NoStop}%
\bibitem [{\citenamefont {Adelberger}\ \emph {et~al.}(2009)\citenamefont
  {Adelberger}, \citenamefont {Gundlach}, \citenamefont {Heckel}, \citenamefont
  {Hoedl},\ and\ \citenamefont {Schlamminger}}]{Adelberger:2009zz}%
  \BibitemOpen
  \bibfield  {author} {\bibinfo {author} {\bibfnamefont {E.~G.}\ \bibnamefont
  {Adelberger}}, \bibinfo {author} {\bibfnamefont {J.~H.}\ \bibnamefont
  {Gundlach}}, \bibinfo {author} {\bibfnamefont {B.~R.}\ \bibnamefont
  {Heckel}}, \bibinfo {author} {\bibfnamefont {S.}~\bibnamefont {Hoedl}},\ and\
  \bibinfo {author} {\bibfnamefont {S.}~\bibnamefont {Schlamminger}},\
  }\bibfield  {title} {\bibinfo {title} {{Torsion balance experiments: A
  low-energy frontier of particle physics}},\ }\href
  {https://doi.org/10.1016/j.ppnp.2008.08.002} {\bibfield  {journal} {\bibinfo
  {journal} {Prog. Part. Nucl. Phys.}\ }\textbf {\bibinfo {volume} {62}},\
  \bibinfo {pages} {102} (\bibinfo {year} {2009})}\BibitemShut {NoStop}%
\bibitem [{\citenamefont {Salumbides}\ \emph {et~al.}(2014)\citenamefont
  {Salumbides}, \citenamefont {Ubachs},\ and\ \citenamefont
  {Korobov}}]{Salumbides:2013dua}%
  \BibitemOpen
  \bibfield  {author} {\bibinfo {author} {\bibfnamefont {E.~J.}\ \bibnamefont
  {Salumbides}}, \bibinfo {author} {\bibfnamefont {W.}~\bibnamefont {Ubachs}},\
  and\ \bibinfo {author} {\bibfnamefont {V.~I.}\ \bibnamefont {Korobov}},\
  }\bibfield  {title} {\bibinfo {title} {{Bounds on fifth forces at the
  sub-Angstrom length scale}},\ }\href
  {https://doi.org/10.1016/j.jms.2014.04.003} {\bibfield  {journal} {\bibinfo
  {journal} {J. Molec. Spectrosc.}\ }\textbf {\bibinfo {volume} {300}},\
  \bibinfo {pages} {65} (\bibinfo {year} {2014})},\ \Eprint
  {https://arxiv.org/abs/1308.1711} {arXiv:1308.1711 [hep-ph]} \BibitemShut
  {NoStop}%
\bibitem [{\citenamefont {Baryakhtar}\ \emph {et~al.}(2017)\citenamefont
  {Baryakhtar}, \citenamefont {Lasenby},\ and\ \citenamefont
  {Teo}}]{Baryakhtar:2017ngi}%
  \BibitemOpen
  \bibfield  {author} {\bibinfo {author} {\bibfnamefont {M.}~\bibnamefont
  {Baryakhtar}}, \bibinfo {author} {\bibfnamefont {R.}~\bibnamefont
  {Lasenby}},\ and\ \bibinfo {author} {\bibfnamefont {M.}~\bibnamefont {Teo}},\
  }\bibfield  {title} {\bibinfo {title} {{Black Hole Superradiance Signatures
  of Ultralight Vectors}},\ }\href {https://doi.org/10.1103/PhysRevD.96.035019}
  {\bibfield  {journal} {\bibinfo  {journal} {Phys. Rev. D}\ }\textbf {\bibinfo
  {volume} {96}},\ \bibinfo {pages} {035019} (\bibinfo {year} {2017})},\
  \Eprint {https://arxiv.org/abs/1704.05081} {arXiv:1704.05081 [hep-ph]}
  \BibitemShut {NoStop}%
\bibitem [{\citenamefont {Poddar}\ \emph {et~al.}(2019)\citenamefont {Poddar},
  \citenamefont {Mohanty},\ and\ \citenamefont {Jana}}]{KumarPoddar:2019ceq}%
  \BibitemOpen
  \bibfield  {author} {\bibinfo {author} {\bibfnamefont {T.~K.}\ \bibnamefont
  {Poddar}}, \bibinfo {author} {\bibfnamefont {S.}~\bibnamefont {Mohanty}},\
  and\ \bibinfo {author} {\bibfnamefont {S.}~\bibnamefont {Jana}},\ }\bibfield
  {title} {\bibinfo {title} {{Vector gauge boson radiation from compact binary
  systems in a gauged $L_\mu-L_\tau$ scenario}},\ }\href
  {https://doi.org/10.1103/PhysRevD.100.123023} {\bibfield  {journal} {\bibinfo
   {journal} {Phys. Rev. D}\ }\textbf {\bibinfo {volume} {100}},\ \bibinfo
  {pages} {123023} (\bibinfo {year} {2019})},\ \Eprint
  {https://arxiv.org/abs/1908.09732} {arXiv:1908.09732 [hep-ph]} \BibitemShut
  {NoStop}%
\bibitem [{\citenamefont {Poddar}\ \emph {et~al.}(2021)\citenamefont {Poddar},
  \citenamefont {Mohanty},\ and\ \citenamefont {Jana}}]{KumarPoddar:2020kdz}%
  \BibitemOpen
  \bibfield  {author} {\bibinfo {author} {\bibfnamefont {T.~K.}\ \bibnamefont
  {Poddar}}, \bibinfo {author} {\bibfnamefont {S.}~\bibnamefont {Mohanty}},\
  and\ \bibinfo {author} {\bibfnamefont {S.}~\bibnamefont {Jana}},\ }\bibfield
  {title} {\bibinfo {title} {{Constraints on long range force from perihelion
  precession of planets in a gauged $L_e-L_{\mu,\tau}$ scenario}},\ }\href
  {https://doi.org/10.1140/epjc/s10052-021-09078-9} {\bibfield  {journal}
  {\bibinfo  {journal} {Eur. Phys. J. C}\ }\textbf {\bibinfo {volume} {81}},\
  \bibinfo {pages} {286} (\bibinfo {year} {2021})},\ \Eprint
  {https://arxiv.org/abs/2002.02935} {arXiv:2002.02935 [hep-ph]} \BibitemShut
  {NoStop}%
\bibitem [{\citenamefont {Honda}\ \emph {et~al.}(2015)\citenamefont {Honda},
  \citenamefont {Sajjad~Athar}, \citenamefont {Kajita}, \citenamefont
  {Kasahara},\ and\ \citenamefont {Midorikawa}}]{Honda:2015fha}%
  \BibitemOpen
  \bibfield  {author} {\bibinfo {author} {\bibfnamefont {M.}~\bibnamefont
  {Honda}}, \bibinfo {author} {\bibfnamefont {M.}~\bibnamefont {Sajjad~Athar}},
  \bibinfo {author} {\bibfnamefont {T.}~\bibnamefont {Kajita}}, \bibinfo
  {author} {\bibfnamefont {K.}~\bibnamefont {Kasahara}},\ and\ \bibinfo
  {author} {\bibfnamefont {S.}~\bibnamefont {Midorikawa}},\ }\bibfield  {title}
  {\bibinfo {title} {{Atmospheric neutrino flux calculation using the
  NRLMSISE-00 atmospheric model}},\ }\href
  {https://doi.org/10.1103/PhysRevD.92.023004} {\bibfield  {journal} {\bibinfo
  {journal} {Phys. Rev. D}\ }\textbf {\bibinfo {volume} {92}},\ \bibinfo
  {pages} {023004} (\bibinfo {year} {2015})},\ \Eprint
  {https://arxiv.org/abs/1502.03916} {arXiv:1502.03916 [astro-ph.HE]}
  \BibitemShut {NoStop}%
\bibitem [{\citenamefont {Abbasi}\ \emph
  {et~al.}(2025{\natexlab{a}})\citenamefont {Abbasi} \emph
  {et~al.}}]{DVN_B4RITM_2025}%
  \BibitemOpen
  \bibfield  {author} {\bibinfo {author} {\bibfnamefont {R.}~\bibnamefont
  {Abbasi}} \emph {et~al.} (\bibinfo {collaboration} {IceCube}),\ }\href
  {https://doi.org/10.7910/DVN/B4RITM} {\bibinfo {title} {{Replication Data
  for: Measurement of atmospheric neutrino mixing with improved IceCube
  DeepCore calibration and data processing}}} (\bibinfo {year}
  {2025}{\natexlab{a}})\BibitemShut {NoStop}%
\bibitem [{\citenamefont {Langacker}(2009)}]{Langacker:2008yv}%
  \BibitemOpen
  \bibfield  {author} {\bibinfo {author} {\bibfnamefont {P.}~\bibnamefont
  {Langacker}},\ }\bibfield  {title} {\bibinfo {title} {{The Physics of Heavy
  $Z^\prime$ Gauge Bosons}},\ }\href
  {https://doi.org/10.1103/RevModPhys.81.1199} {\bibfield  {journal} {\bibinfo
  {journal} {Rev. Mod. Phys.}\ }\textbf {\bibinfo {volume} {81}},\ \bibinfo
  {pages} {1199} (\bibinfo {year} {2009})},\ \Eprint
  {https://arxiv.org/abs/0801.1345} {arXiv:0801.1345 [hep-ph]} \BibitemShut
  {NoStop}%
\bibitem [{\citenamefont {He}\ \emph {et~al.}(1991{\natexlab{a}})\citenamefont
  {He}, \citenamefont {Joshi}, \citenamefont {Lew},\ and\ \citenamefont
  {Volkas}}]{He:1990pn}%
  \BibitemOpen
  \bibfield  {author} {\bibinfo {author} {\bibfnamefont {X.~G.}\ \bibnamefont
  {He}}, \bibinfo {author} {\bibfnamefont {G.~C.}\ \bibnamefont {Joshi}},
  \bibinfo {author} {\bibfnamefont {H.}~\bibnamefont {Lew}},\ and\ \bibinfo
  {author} {\bibfnamefont {R.~R.}\ \bibnamefont {Volkas}},\ }\bibfield  {title}
  {\bibinfo {title} {{NEW Z-prime PHENOMENOLOGY}},\ }\href
  {https://doi.org/10.1103/PhysRevD.43.R22} {\bibfield  {journal} {\bibinfo
  {journal} {Phys. Rev. D}\ }\textbf {\bibinfo {volume} {43}},\ \bibinfo
  {pages} {22} (\bibinfo {year} {1991}{\natexlab{a}})}\BibitemShut {NoStop}%
\bibitem [{\citenamefont {Foot}(1991)}]{Foot:1990mn}%
  \BibitemOpen
  \bibfield  {author} {\bibinfo {author} {\bibfnamefont {R.}~\bibnamefont
  {Foot}},\ }\bibfield  {title} {\bibinfo {title} {{New Physics From Electric
  Charge Quantization?}},\ }\href {https://doi.org/10.1142/S0217732391000543}
  {\bibfield  {journal} {\bibinfo  {journal} {Mod. Phys. Lett. A}\ }\textbf
  {\bibinfo {volume} {6}},\ \bibinfo {pages} {527} (\bibinfo {year}
  {1991})}\BibitemShut {NoStop}%
\bibitem [{\citenamefont {He}\ \emph {et~al.}(1991{\natexlab{b}})\citenamefont
  {He}, \citenamefont {Joshi}, \citenamefont {Lew},\ and\ \citenamefont
  {Volkas}}]{He:1991qd}%
  \BibitemOpen
  \bibfield  {author} {\bibinfo {author} {\bibfnamefont {X.-G.}\ \bibnamefont
  {He}}, \bibinfo {author} {\bibfnamefont {G.~C.}\ \bibnamefont {Joshi}},
  \bibinfo {author} {\bibfnamefont {H.}~\bibnamefont {Lew}},\ and\ \bibinfo
  {author} {\bibfnamefont {R.~R.}\ \bibnamefont {Volkas}},\ }\bibfield  {title}
  {\bibinfo {title} {{Simplest Z-prime model}},\ }\href
  {https://doi.org/10.1103/PhysRevD.44.2118} {\bibfield  {journal} {\bibinfo
  {journal} {Phys. Rev. D}\ }\textbf {\bibinfo {volume} {44}},\ \bibinfo
  {pages} {2118} (\bibinfo {year} {1991}{\natexlab{b}})}\BibitemShut {NoStop}%
\bibitem [{\citenamefont {Foot}\ \emph {et~al.}(1994)\citenamefont {Foot},
  \citenamefont {He}, \citenamefont {Lew},\ and\ \citenamefont
  {Volkas}}]{Foot:1994vd}%
  \BibitemOpen
  \bibfield  {author} {\bibinfo {author} {\bibfnamefont {R.}~\bibnamefont
  {Foot}}, \bibinfo {author} {\bibfnamefont {X.~G.}\ \bibnamefont {He}},
  \bibinfo {author} {\bibfnamefont {H.}~\bibnamefont {Lew}},\ and\ \bibinfo
  {author} {\bibfnamefont {R.~R.}\ \bibnamefont {Volkas}},\ }\bibfield  {title}
  {\bibinfo {title} {{Model for a light Z-prime boson}},\ }\href
  {https://doi.org/10.1103/PhysRevD.50.4571} {\bibfield  {journal} {\bibinfo
  {journal} {Phys. Rev. D}\ }\textbf {\bibinfo {volume} {50}},\ \bibinfo
  {pages} {4571} (\bibinfo {year} {1994})},\ \Eprint
  {https://arxiv.org/abs/hep-ph/9401250} {arXiv:hep-ph/9401250} \BibitemShut
  {NoStop}%
\bibitem [{\citenamefont {Dutta}\ \emph {et~al.}(1994)\citenamefont {Dutta},
  \citenamefont {Joshipura},\ and\ \citenamefont {Vijaykumar}}]{Dutta:1994dx}%
  \BibitemOpen
  \bibfield  {author} {\bibinfo {author} {\bibfnamefont {G.}~\bibnamefont
  {Dutta}}, \bibinfo {author} {\bibfnamefont {A.~S.}\ \bibnamefont
  {Joshipura}},\ and\ \bibinfo {author} {\bibfnamefont {K.~B.}\ \bibnamefont
  {Vijaykumar}},\ }\bibfield  {title} {\bibinfo {title} {{Leptonic flavor
  violations in the presence of an extra Z}},\ }\href
  {https://doi.org/10.1103/PhysRevD.50.2109} {\bibfield  {journal} {\bibinfo
  {journal} {Phys. Rev. D}\ }\textbf {\bibinfo {volume} {50}},\ \bibinfo
  {pages} {2109} (\bibinfo {year} {1994})},\ \Eprint
  {https://arxiv.org/abs/hep-ph/9405292} {arXiv:hep-ph/9405292} \BibitemShut
  {NoStop}%
\bibitem [{\citenamefont {Wolfenstein}(1978)}]{Wolfenstein:1977ue}%
  \BibitemOpen
  \bibfield  {author} {\bibinfo {author} {\bibfnamefont {L.}~\bibnamefont
  {Wolfenstein}},\ }\bibfield  {title} {\bibinfo {title} {{Neutrino
  Oscillations in Matter}},\ }\href {https://doi.org/10.1103/PhysRevD.17.2369}
  {\bibfield  {journal} {\bibinfo  {journal} {Phys. Rev. D}\ }\textbf {\bibinfo
  {volume} {17}},\ \bibinfo {pages} {2369} (\bibinfo {year}
  {1978})}\BibitemShut {NoStop}%
\bibitem [{\citenamefont {Valle}(1987)}]{Valle:1987gv}%
  \BibitemOpen
  \bibfield  {author} {\bibinfo {author} {\bibfnamefont {J.~W.~F.}\
  \bibnamefont {Valle}},\ }\bibfield  {title} {\bibinfo {title} {{Resonant
  Oscillations of Massless Neutrinos in Matter}},\ }\href
  {https://doi.org/10.1016/0370-2693(87)90947-6} {\bibfield  {journal}
  {\bibinfo  {journal} {Phys. Lett. B}\ }\textbf {\bibinfo {volume} {199}},\
  \bibinfo {pages} {432} (\bibinfo {year} {1987})}\BibitemShut {NoStop}%
\bibitem [{\citenamefont {Guzzo}\ \emph {et~al.}(1991)\citenamefont {Guzzo},
  \citenamefont {Masiero},\ and\ \citenamefont {Petcov}}]{Guzzo:1991hi}%
  \BibitemOpen
  \bibfield  {author} {\bibinfo {author} {\bibfnamefont {M.~M.}\ \bibnamefont
  {Guzzo}}, \bibinfo {author} {\bibfnamefont {A.}~\bibnamefont {Masiero}},\
  and\ \bibinfo {author} {\bibfnamefont {S.~T.}\ \bibnamefont {Petcov}},\
  }\bibfield  {title} {\bibinfo {title} {{On the MSW effect with massless
  neutrinos and no mixing in the vacuum}},\ }\href
  {https://doi.org/10.1016/0370-2693(91)90984-X} {\bibfield  {journal}
  {\bibinfo  {journal} {Phys. Lett. B}\ }\textbf {\bibinfo {volume} {260}},\
  \bibinfo {pages} {154} (\bibinfo {year} {1991})}\BibitemShut {NoStop}%
\bibitem [{\citenamefont {Mitsuka}\ \emph {et~al.}(2011)\citenamefont {Mitsuka}
  \emph {et~al.}}]{Super-Kamiokande:2011dam}%
  \BibitemOpen
  \bibfield  {author} {\bibinfo {author} {\bibfnamefont {G.}~\bibnamefont
  {Mitsuka}} \emph {et~al.} (\bibinfo {collaboration} {Super-Kamiokande}),\
  }\bibfield  {title} {\bibinfo {title} {{Study of Non-Standard Neutrino
  Interactions with Atmospheric Neutrino Data in Super-Kamiokande I and II}},\
  }\href {https://doi.org/10.1103/PhysRevD.84.113008} {\bibfield  {journal}
  {\bibinfo  {journal} {Phys. Rev. D}\ }\textbf {\bibinfo {volume} {84}},\
  \bibinfo {pages} {113008} (\bibinfo {year} {2011})},\ \Eprint
  {https://arxiv.org/abs/1109.1889} {arXiv:1109.1889 [hep-ex]} \BibitemShut
  {NoStop}%
\bibitem [{\citenamefont {Ohlsson}(2013)}]{Ohlsson:2012kf}%
  \BibitemOpen
  \bibfield  {author} {\bibinfo {author} {\bibfnamefont {T.}~\bibnamefont
  {Ohlsson}},\ }\bibfield  {title} {\bibinfo {title} {{Status of non-standard
  neutrino interactions}},\ }\href
  {https://doi.org/10.1088/0034-4885/76/4/044201} {\bibfield  {journal}
  {\bibinfo  {journal} {Rept. Prog. Phys.}\ }\textbf {\bibinfo {volume} {76}},\
  \bibinfo {pages} {044201} (\bibinfo {year} {2013})},\ \Eprint
  {https://arxiv.org/abs/1209.2710} {arXiv:1209.2710 [hep-ph]} \BibitemShut
  {NoStop}%
\bibitem [{\citenamefont {Gonzalez-Garcia}\ and\ \citenamefont
  {Maltoni}(2013)}]{Gonzalez-Garcia:2013usa}%
  \BibitemOpen
  \bibfield  {author} {\bibinfo {author} {\bibfnamefont {M.~C.}\ \bibnamefont
  {Gonzalez-Garcia}}\ and\ \bibinfo {author} {\bibfnamefont {M.}~\bibnamefont
  {Maltoni}},\ }\bibfield  {title} {\bibinfo {title} {{Determination of matter
  potential from global analysis of neutrino oscillation data}},\ }\href
  {https://doi.org/10.1007/JHEP09(2013)152} {\bibfield  {journal} {\bibinfo
  {journal} {JHEP}\ }\textbf {\bibinfo {volume} {09}},\ \bibinfo {pages}
  {152}},\ \Eprint {https://arxiv.org/abs/1307.3092} {arXiv:1307.3092 [hep-ph]}
  \BibitemShut {NoStop}%
\bibitem [{\citenamefont {Arkani-Hamed}\ \emph {et~al.}(2007)\citenamefont
  {Arkani-Hamed}, \citenamefont {Motl}, \citenamefont {Nicolis},\ and\
  \citenamefont {Vafa}}]{Arkani-Hamed:2006emk}%
  \BibitemOpen
  \bibfield  {author} {\bibinfo {author} {\bibfnamefont {N.}~\bibnamefont
  {Arkani-Hamed}}, \bibinfo {author} {\bibfnamefont {L.}~\bibnamefont {Motl}},
  \bibinfo {author} {\bibfnamefont {A.}~\bibnamefont {Nicolis}},\ and\ \bibinfo
  {author} {\bibfnamefont {C.}~\bibnamefont {Vafa}},\ }\bibfield  {title}
  {\bibinfo {title} {{The String landscape, black holes and gravity as the
  weakest force}},\ }\href {https://doi.org/10.1088/1126-6708/2007/06/060}
  {\bibfield  {journal} {\bibinfo  {journal} {JHEP}\ }\textbf {\bibinfo
  {volume} {06}},\ \bibinfo {pages} {060}},\ \Eprint
  {https://arxiv.org/abs/hep-th/0601001} {arXiv:hep-th/0601001} \BibitemShut
  {NoStop}%
\bibitem [{\citenamefont {Maki}\ \emph {et~al.}(1962)\citenamefont {Maki},
  \citenamefont {Nakagawa},\ and\ \citenamefont {Sakata}}]{Maki:1962mu}%
  \BibitemOpen
  \bibfield  {author} {\bibinfo {author} {\bibfnamefont {Z.}~\bibnamefont
  {Maki}}, \bibinfo {author} {\bibfnamefont {M.}~\bibnamefont {Nakagawa}},\
  and\ \bibinfo {author} {\bibfnamefont {S.}~\bibnamefont {Sakata}},\
  }\bibfield  {title} {\bibinfo {title} {{Remarks on the unified model of
  elementary particles}},\ }\href {https://doi.org/10.1143/PTP.28.870}
  {\bibfield  {journal} {\bibinfo  {journal} {Prog. Theor. Phys.}\ }\textbf
  {\bibinfo {volume} {28}},\ \bibinfo {pages} {870} (\bibinfo {year}
  {1962})}\BibitemShut {NoStop}%
\bibitem [{\citenamefont {Pontecorvo}(1967)}]{Pontecorvo:1967fh}%
  \BibitemOpen
  \bibfield  {author} {\bibinfo {author} {\bibfnamefont {B.}~\bibnamefont
  {Pontecorvo}},\ }\bibfield  {title} {\bibinfo {title} {{Neutrino Experiments
  and the Problem of Conservation of Leptonic Charge}},\ }\href@noop {}
  {\bibfield  {journal} {\bibinfo  {journal} {Zh. Eksp. Teor. Fiz.}\ }\textbf
  {\bibinfo {volume} {53}},\ \bibinfo {pages} {1717} (\bibinfo {year}
  {1967})}\BibitemShut {NoStop}%
\bibitem [{\citenamefont {Dziewonski}\ and\ \citenamefont
  {Anderson}(1981)}]{Dziewonski:1981xy}%
  \BibitemOpen
  \bibfield  {author} {\bibinfo {author} {\bibfnamefont {A.~M.}\ \bibnamefont
  {Dziewonski}}\ and\ \bibinfo {author} {\bibfnamefont {D.~L.}\ \bibnamefont
  {Anderson}},\ }\bibfield  {title} {\bibinfo {title} {{Preliminary reference
  earth model}},\ }\href {https://doi.org/10.1016/0031-9201(81)90046-7}
  {\bibfield  {journal} {\bibinfo  {journal} {Phys. Earth Planet. Interiors}\
  }\textbf {\bibinfo {volume} {25}},\ \bibinfo {pages} {297} (\bibinfo {year}
  {1981})}\BibitemShut {NoStop}%
\bibitem [{\citenamefont {Aartsen}\ \emph
  {et~al.}(2017{\natexlab{a}})\citenamefont {Aartsen} \emph
  {et~al.}}]{IceCube:2016zyt}%
  \BibitemOpen
  \bibfield  {author} {\bibinfo {author} {\bibfnamefont {M.~G.}\ \bibnamefont
  {Aartsen}} \emph {et~al.} (\bibinfo {collaboration} {IceCube}),\ }\bibfield
  {title} {\bibinfo {title} {{The IceCube Neutrino Observatory: Instrumentation
  and Online Systems}},\ }\href
  {https://doi.org/10.1088/1748-0221/12/03/P03012} {\bibfield  {journal}
  {\bibinfo  {journal} {JINST}\ }\textbf {\bibinfo {volume} {12}}\bibfield
  {number} {\bibinfo  {number} { (03)},\ \bibinfo {pages} {P03012}},\ }\bibinfo
  {note} {[Erratum: JINST 19, E05001 (2024)]},\ \Eprint
  {https://arxiv.org/abs/1612.05093} {arXiv:1612.05093 [astro-ph.IM]}
  \BibitemShut {NoStop}%
\bibitem [{\citenamefont {Abbasi}\ \emph {et~al.}(2010)\citenamefont {Abbasi}
  \emph {et~al.}}]{IceCube:2010dpc}%
  \BibitemOpen
  \bibfield  {author} {\bibinfo {author} {\bibfnamefont {R.}~\bibnamefont
  {Abbasi}} \emph {et~al.} (\bibinfo {collaboration} {IceCube}),\ }\bibfield
  {title} {\bibinfo {title} {{Calibration and Characterization of the IceCube
  Photomultiplier Tube}},\ }\href {https://doi.org/10.1016/j.nima.2010.03.102}
  {\bibfield  {journal} {\bibinfo  {journal} {Nucl. Instrum. Meth. A}\ }\textbf
  {\bibinfo {volume} {618}},\ \bibinfo {pages} {139} (\bibinfo {year}
  {2010})},\ \Eprint {https://arxiv.org/abs/1002.2442} {arXiv:1002.2442
  [astro-ph.IM]} \BibitemShut {NoStop}%
\bibitem [{\citenamefont {Abbasi}\ \emph {et~al.}(2012)\citenamefont {Abbasi}
  \emph {et~al.}}]{IceCube:2011ucd}%
  \BibitemOpen
  \bibfield  {author} {\bibinfo {author} {\bibfnamefont {R.}~\bibnamefont
  {Abbasi}} \emph {et~al.} (\bibinfo {collaboration} {IceCube}),\ }\bibfield
  {title} {\bibinfo {title} {{The Design and Performance of IceCube
  DeepCore}},\ }\href {https://doi.org/10.1016/j.astropartphys.2012.01.004}
  {\bibfield  {journal} {\bibinfo  {journal} {Astropart. Phys.}\ }\textbf
  {\bibinfo {volume} {35}},\ \bibinfo {pages} {615} (\bibinfo {year} {2012})},\
  \Eprint {https://arxiv.org/abs/1109.6096} {arXiv:1109.6096 [astro-ph.IM]}
  \BibitemShut {NoStop}%
\bibitem [{\citenamefont {Aartsen}\ \emph
  {et~al.}(2018{\natexlab{a}})\citenamefont {Aartsen} \emph
  {et~al.}}]{IceCube:2017lak}%
  \BibitemOpen
  \bibfield  {author} {\bibinfo {author} {\bibfnamefont {M.~G.}\ \bibnamefont
  {Aartsen}} \emph {et~al.} (\bibinfo {collaboration} {IceCube}),\ }\bibfield
  {title} {\bibinfo {title} {{Measurement of Atmospheric Neutrino Oscillations
  at 6{\textendash}56 GeV with IceCube DeepCore}},\ }\href
  {https://doi.org/10.1103/PhysRevLett.120.071801} {\bibfield  {journal}
  {\bibinfo  {journal} {Phys. Rev. Lett.}\ }\textbf {\bibinfo {volume} {120}},\
  \bibinfo {pages} {071801} (\bibinfo {year} {2018}{\natexlab{a}})},\ \Eprint
  {https://arxiv.org/abs/1707.07081} {arXiv:1707.07081 [hep-ex]} \BibitemShut
  {NoStop}%
\bibitem [{\citenamefont {Abbasi}\ \emph {et~al.}(2023)\citenamefont {Abbasi}
  \emph {et~al.}}]{IceCubeCollaboration:2023wtb}%
  \BibitemOpen
  \bibfield  {author} {\bibinfo {author} {\bibfnamefont {R.}~\bibnamefont
  {Abbasi}} \emph {et~al.} (\bibinfo {collaboration} {IceCube}),\ }\bibfield
  {title} {\bibinfo {title} {{Measurement of atmospheric neutrino mixing with
  improved IceCube DeepCore calibration and data processing}},\ }\href
  {https://doi.org/10.1103/PhysRevD.108.012014} {\bibfield  {journal} {\bibinfo
   {journal} {Phys. Rev. D}\ }\textbf {\bibinfo {volume} {108}},\ \bibinfo
  {pages} {012014} (\bibinfo {year} {2023})},\ \Eprint
  {https://arxiv.org/abs/2304.12236} {arXiv:2304.12236 [hep-ex]} \BibitemShut
  {NoStop}%
\bibitem [{\citenamefont {Abbasi}\ \emph
  {et~al.}(2025{\natexlab{b}})\citenamefont {Abbasi} \emph
  {et~al.}}]{IceCubeCollaboration:2024ssx}%
  \BibitemOpen
  \bibfield  {author} {\bibinfo {author} {\bibfnamefont {R.}~\bibnamefont
  {Abbasi}} \emph {et~al.} (\bibinfo {collaboration} {IceCube}),\ }\bibfield
  {title} {\bibinfo {title} {{Measurement of Atmospheric Neutrino Oscillation
  Parameters Using Convolutional Neural Networks with 9.3 Years of Data in
  IceCube DeepCore}},\ }\href {https://doi.org/10.1103/PhysRevLett.134.091801}
  {\bibfield  {journal} {\bibinfo  {journal} {Phys. Rev. Lett.}\ }\textbf
  {\bibinfo {volume} {134}},\ \bibinfo {pages} {091801} (\bibinfo {year}
  {2025}{\natexlab{b}})},\ \Eprint {https://arxiv.org/abs/2405.02163}
  {arXiv:2405.02163 [hep-ex]} \BibitemShut {NoStop}%
\bibitem [{\citenamefont {Aartsen}\ \emph {et~al.}(2019)\citenamefont {Aartsen}
  \emph {et~al.}}]{IceCube:2019dqi}%
  \BibitemOpen
  \bibfield  {author} {\bibinfo {author} {\bibfnamefont {M.~G.}\ \bibnamefont
  {Aartsen}} \emph {et~al.} (\bibinfo {collaboration} {IceCube}),\ }\bibfield
  {title} {\bibinfo {title} {{Measurement of Atmospheric Tau Neutrino
  Appearance with IceCube DeepCore}},\ }\href
  {https://doi.org/10.1103/PhysRevD.99.032007} {\bibfield  {journal} {\bibinfo
  {journal} {Phys. Rev. D}\ }\textbf {\bibinfo {volume} {99}},\ \bibinfo
  {pages} {032007} (\bibinfo {year} {2019})},\ \Eprint
  {https://arxiv.org/abs/1901.05366} {arXiv:1901.05366 [hep-ex]} \BibitemShut
  {NoStop}%
\bibitem [{\citenamefont {Aartsen}\ \emph
  {et~al.}(2017{\natexlab{b}})\citenamefont {Aartsen} \emph
  {et~al.}}]{IceCube:2017ivd}%
  \BibitemOpen
  \bibfield  {author} {\bibinfo {author} {\bibfnamefont {M.~G.}\ \bibnamefont
  {Aartsen}} \emph {et~al.} (\bibinfo {collaboration} {IceCube}),\ }\bibfield
  {title} {\bibinfo {title} {{Search for sterile neutrino mixing using three
  years of IceCube DeepCore data}},\ }\href
  {https://doi.org/10.1103/PhysRevD.95.112002} {\bibfield  {journal} {\bibinfo
  {journal} {Phys. Rev. D}\ }\textbf {\bibinfo {volume} {95}},\ \bibinfo
  {pages} {112002} (\bibinfo {year} {2017}{\natexlab{b}})},\ \Eprint
  {https://arxiv.org/abs/1702.05160} {arXiv:1702.05160 [hep-ex]} \BibitemShut
  {NoStop}%
\bibitem [{\citenamefont {Aartsen}\ \emph
  {et~al.}(2020{\natexlab{a}})\citenamefont {Aartsen} \emph
  {et~al.}}]{IceCube:2020phf}%
  \BibitemOpen
  \bibfield  {author} {\bibinfo {author} {\bibfnamefont {M.~G.}\ \bibnamefont
  {Aartsen}} \emph {et~al.} (\bibinfo {collaboration} {IceCube}),\ }\bibfield
  {title} {\bibinfo {title} {{eV-Scale Sterile Neutrino Search Using Eight
  Years of Atmospheric Muon Neutrino Data from the IceCube Neutrino
  Observatory}},\ }\href {https://doi.org/10.1103/PhysRevLett.125.141801}
  {\bibfield  {journal} {\bibinfo  {journal} {Phys. Rev. Lett.}\ }\textbf
  {\bibinfo {volume} {125}},\ \bibinfo {pages} {141801} (\bibinfo {year}
  {2020}{\natexlab{a}})},\ \Eprint {https://arxiv.org/abs/2005.12942}
  {arXiv:2005.12942 [hep-ex]} \BibitemShut {NoStop}%
\bibitem [{\citenamefont {Aartsen}\ \emph
  {et~al.}(2020{\natexlab{b}})\citenamefont {Aartsen} \emph
  {et~al.}}]{IceCube:2020tka}%
  \BibitemOpen
  \bibfield  {author} {\bibinfo {author} {\bibfnamefont {M.~G.}\ \bibnamefont
  {Aartsen}} \emph {et~al.} (\bibinfo {collaboration} {IceCube}),\ }\bibfield
  {title} {\bibinfo {title} {{Searching for eV-scale sterile neutrinos with
  eight years of atmospheric neutrinos at the IceCube Neutrino Telescope}},\
  }\href {https://doi.org/10.1103/PhysRevD.102.052009} {\bibfield  {journal}
  {\bibinfo  {journal} {Phys. Rev. D}\ }\textbf {\bibinfo {volume} {102}},\
  \bibinfo {pages} {052009} (\bibinfo {year} {2020}{\natexlab{b}})},\ \Eprint
  {https://arxiv.org/abs/2005.12943} {arXiv:2005.12943 [hep-ex]} \BibitemShut
  {NoStop}%
\bibitem [{\citenamefont {Abbasi}\ \emph
  {et~al.}(2024{\natexlab{a}})\citenamefont {Abbasi} \emph
  {et~al.}}]{IceCubeCollaboration:2024nle}%
  \BibitemOpen
  \bibfield  {author} {\bibinfo {author} {\bibfnamefont {R.}~\bibnamefont
  {Abbasi}} \emph {et~al.} (\bibinfo {collaboration} {IceCube}),\ }\bibfield
  {title} {\bibinfo {title} {{Search for an eV-Scale Sterile Neutrino Using
  Improved High-Energy $\nu_\mu$ Event Reconstruction in IceCube}},\ }\href
  {https://doi.org/10.1103/PhysRevLett.133.201804} {\bibfield  {journal}
  {\bibinfo  {journal} {Phys. Rev. Lett.}\ }\textbf {\bibinfo {volume} {133}},\
  \bibinfo {pages} {201804} (\bibinfo {year} {2024}{\natexlab{a}})},\ \Eprint
  {https://arxiv.org/abs/2405.08070} {arXiv:2405.08070 [hep-ex]} \BibitemShut
  {NoStop}%
\bibitem [{\citenamefont {Abbasi}\ \emph
  {et~al.}(2024{\natexlab{b}})\citenamefont {Abbasi} \emph
  {et~al.}}]{IceCube:2024pky}%
  \BibitemOpen
  \bibfield  {author} {\bibinfo {author} {\bibfnamefont {R.}~\bibnamefont
  {Abbasi}} \emph {et~al.} (\bibinfo {collaboration} {IceCube}),\ }\bibfield
  {title} {\bibinfo {title} {{Exploration of mass splitting and muon/tau mixing
  parameters for an eV-scale sterile neutrino with IceCube}},\ }\href
  {https://doi.org/10.1016/j.physletb.2024.139077} {\bibfield  {journal}
  {\bibinfo  {journal} {Phys. Lett. B}\ }\textbf {\bibinfo {volume} {858}},\
  \bibinfo {pages} {139077} (\bibinfo {year} {2024}{\natexlab{b}})},\ \Eprint
  {https://arxiv.org/abs/2406.00905} {arXiv:2406.00905 [hep-ex]} \BibitemShut
  {NoStop}%
\bibitem [{\citenamefont {Abbasi}\ \emph
  {et~al.}(2024{\natexlab{c}})\citenamefont {Abbasi} \emph
  {et~al.}}]{IceCube:2024dlz}%
  \BibitemOpen
  \bibfield  {author} {\bibinfo {author} {\bibfnamefont {R.}~\bibnamefont
  {Abbasi}} \emph {et~al.} (\bibinfo {collaboration} {IceCube}),\ }\bibfield
  {title} {\bibinfo {title} {{Search for a light sterile neutrino with
  7.5~years of IceCube DeepCore data}},\ }\href
  {https://doi.org/10.1103/PhysRevD.110.072007} {\bibfield  {journal} {\bibinfo
   {journal} {Phys. Rev. D}\ }\textbf {\bibinfo {volume} {110}},\ \bibinfo
  {pages} {072007} (\bibinfo {year} {2024}{\natexlab{c}})},\ \Eprint
  {https://arxiv.org/abs/2407.01314} {arXiv:2407.01314 [hep-ex]} \BibitemShut
  {NoStop}%
\bibitem [{\citenamefont {Aartsen}\ \emph
  {et~al.}(2018{\natexlab{b}})\citenamefont {Aartsen} \emph
  {et~al.}}]{IceCube:2017zcu}%
  \BibitemOpen
  \bibfield  {author} {\bibinfo {author} {\bibfnamefont {M.~G.}\ \bibnamefont
  {Aartsen}} \emph {et~al.} (\bibinfo {collaboration} {IceCube}),\ }\bibfield
  {title} {\bibinfo {title} {{Search for Nonstandard Neutrino Interactions with
  IceCube DeepCore}},\ }\href {https://doi.org/10.1103/PhysRevD.97.072009}
  {\bibfield  {journal} {\bibinfo  {journal} {Phys. Rev. D}\ }\textbf {\bibinfo
  {volume} {97}},\ \bibinfo {pages} {072009} (\bibinfo {year}
  {2018}{\natexlab{b}})},\ \Eprint {https://arxiv.org/abs/1709.07079}
  {arXiv:1709.07079 [hep-ex]} \BibitemShut {NoStop}%
\bibitem [{\citenamefont {Abbasi}\ \emph {et~al.}(2021)\citenamefont {Abbasi}
  \emph {et~al.}}]{IceCubeCollaboration:2021euf}%
  \BibitemOpen
  \bibfield  {author} {\bibinfo {author} {\bibfnamefont {R.}~\bibnamefont
  {Abbasi}} \emph {et~al.} (\bibinfo {collaboration} {IceCube}),\ }\bibfield
  {title} {\bibinfo {title} {{All-flavor constraints on nonstandard neutrino
  interactions and generalized matter potential with three years of IceCube
  DeepCore data}},\ }\href {https://doi.org/10.1103/PhysRevD.104.072006}
  {\bibfield  {journal} {\bibinfo  {journal} {Phys. Rev. D}\ }\textbf {\bibinfo
  {volume} {104}},\ \bibinfo {pages} {072006} (\bibinfo {year} {2021})},\
  \Eprint {https://arxiv.org/abs/2106.07755} {arXiv:2106.07755 [hep-ex]}
  \BibitemShut {NoStop}%
\bibitem [{\citenamefont {Abbasi}\ \emph
  {et~al.}(2025{\natexlab{c}})\citenamefont {Abbasi} \emph
  {et~al.}}]{IceCube:2025kve}%
  \BibitemOpen
  \bibfield  {author} {\bibinfo {author} {\bibfnamefont {R.}~\bibnamefont
  {Abbasi}} \emph {et~al.} (\bibinfo {collaboration} {IceCube}),\ }\bibfield
  {title} {\bibinfo {title} {{Search for Heavy Neutral Leptons with IceCube
  DeepCore}},\ }\href@noop {} {\  (\bibinfo {year} {2025}{\natexlab{c}})},\
  \Eprint {https://arxiv.org/abs/2502.09454} {arXiv:2502.09454 [hep-ex]}
  \BibitemShut {NoStop}%
\bibitem [{\citenamefont {Aartsen}\ \emph
  {et~al.}(2020{\natexlab{c}})\citenamefont {Aartsen} \emph
  {et~al.}}]{IceCube:2020nwx}%
  \BibitemOpen
  \bibfield  {author} {\bibinfo {author} {\bibfnamefont {M.~G.}\ \bibnamefont
  {Aartsen}} \emph {et~al.} (\bibinfo {collaboration} {IceCube}),\ }\bibfield
  {title} {\bibinfo {title} {{In-situ calibration of the single-photoelectron
  charge response of the IceCube photomultiplier tubes}},\ }\href
  {https://doi.org/10.1088/1748-0221/15/06/P06032} {\bibfield  {journal}
  {\bibinfo  {journal} {JINST}\ }\textbf {\bibinfo {volume} {15}}\bibfield
  {number} {\bibinfo  {number} { (06)},\ \bibinfo {pages} {P06032}},\ }\Eprint
  {https://arxiv.org/abs/2002.00997} {arXiv:2002.00997 [physics.ins-det]}
  \BibitemShut {NoStop}%
\bibitem [{\citenamefont {Chirkin}(2013)}]{Chirkin:2013tma}%
  \BibitemOpen
  \bibfield  {author} {\bibinfo {author} {\bibfnamefont {D.}~\bibnamefont
  {Chirkin}} (\bibinfo {collaboration} {IceCube}),\ }\bibfield  {title}
  {\bibinfo {title} {{Photon tracking with GPUs in IceCube}},\ }\href
  {https://doi.org/10.1016/j.nima.2012.11.170} {\bibfield  {journal} {\bibinfo
  {journal} {Nucl. Instrum. Meth. A}\ }\textbf {\bibinfo {volume} {725}},\
  \bibinfo {pages} {141} (\bibinfo {year} {2013})}\BibitemShut {NoStop}%
\bibitem [{\citenamefont {Abbasi}\ \emph {et~al.}(2022)\citenamefont {Abbasi}
  \emph {et~al.}}]{IceCube:2022kff}%
  \BibitemOpen
  \bibfield  {author} {\bibinfo {author} {\bibfnamefont {R.}~\bibnamefont
  {Abbasi}} \emph {et~al.} (\bibinfo {collaboration} {IceCube}),\ }\bibfield
  {title} {\bibinfo {title} {{Low energy event reconstruction in IceCube
  DeepCore}},\ }\href {https://doi.org/10.1140/epjc/s10052-022-10721-2}
  {\bibfield  {journal} {\bibinfo  {journal} {Eur. Phys. J. C}\ }\textbf
  {\bibinfo {volume} {82}},\ \bibinfo {pages} {807} (\bibinfo {year} {2022})},\
  \Eprint {https://arxiv.org/abs/2203.02303} {arXiv:2203.02303 [hep-ex]}
  \BibitemShut {NoStop}%
\bibitem [{\citenamefont {Aartsen}\ \emph
  {et~al.}(2020{\natexlab{d}})\citenamefont {Aartsen} \emph
  {et~al.}}]{IceCube:2018ikn}%
  \BibitemOpen
  \bibfield  {author} {\bibinfo {author} {\bibfnamefont {M.~G.}\ \bibnamefont
  {Aartsen}} \emph {et~al.} (\bibinfo {collaboration} {IceCube}),\ }\bibfield
  {title} {\bibinfo {title} {{Computational techniques for the analysis of
  small signals in high-statistics neutrino oscillation experiments}},\ }\href
  {https://doi.org/10.1016/j.nima.2020.164332} {\bibfield  {journal} {\bibinfo
  {journal} {Nucl. Instrum. Meth. A}\ }\textbf {\bibinfo {volume} {977}},\
  \bibinfo {pages} {164332} (\bibinfo {year} {2020}{\natexlab{d}})},\ \Eprint
  {https://arxiv.org/abs/1803.05390} {arXiv:1803.05390 [physics.data-an]}
  \BibitemShut {NoStop}%
\bibitem [{\citenamefont {Y{\'a}{\~n}ez}(2014)}]{Garza2014Measurement}%
  \BibitemOpen
  \bibfield  {author} {\bibinfo {author} {\bibfnamefont {J.~P.}\ \bibnamefont
  {Y{\'a}{\~n}ez}},\ }\emph {\bibinfo {title} {{Measurement of neutrino
  oscillations in atmospheric neutrinos with the IceCube DeepCore detector}}},\
  \href {https://doi.org/http://dx.doi.org/10.18452/17016} {Ph.D. thesis},\
  \bibinfo  {school} {Humboldt-Universit{\"a}t zu Berlin,
  Mathematisch-Naturwissenschaftliche Fakult{\"a}t I} (\bibinfo {year}
  {2014})\BibitemShut {NoStop}%
\bibitem [{\citenamefont {Terliuk}(2018)}]{AndriiThesis}%
  \BibitemOpen
  \bibfield  {author} {\bibinfo {author} {\bibfnamefont {A.}~\bibnamefont
  {Terliuk}},\ }\emph {\bibinfo {title} {{Measurement of atmospheric neutrino
  oscillations and search for sterile neutrino mixing with IceCube
  DeepCore}}},\ \href {https://doi.org/http://dx.doi.org/10.18452/19304} {Ph.D.
  thesis},\ \bibinfo  {school} {Humboldt-Universit{\"a}t zu Berlin,
  Mathematisch-Naturwissenschaftliche Fakult{\"a}t I} (\bibinfo {year}
  {2018})\BibitemShut {NoStop}%
\bibitem [{\citenamefont {Friedman}(2001)}]{Friedman:2001wbq}%
  \BibitemOpen
  \bibfield  {author} {\bibinfo {author} {\bibfnamefont {J.~H.}\ \bibnamefont
  {Friedman}},\ }\bibfield  {title} {\bibinfo {title} {{Greedy function
  approximation: A gradient boosting machine.}},\ }\href
  {https://doi.org/10.1214/aos/1013203451} {\bibfield  {journal} {\bibinfo
  {journal} {Annals Statist.}\ }\textbf {\bibinfo {volume} {29}},\ \bibinfo
  {pages} {1189} (\bibinfo {year} {2001})}\BibitemShut {NoStop}%
\bibitem [{\citenamefont {Esteban}\ \emph {et~al.}(2020)\citenamefont
  {Esteban}, \citenamefont {Gonzalez-Garcia}, \citenamefont {Maltoni},
  \citenamefont {Schwetz},\ and\ \citenamefont {Zhou}}]{Esteban:2020cvm}%
  \BibitemOpen
  \bibfield  {author} {\bibinfo {author} {\bibfnamefont {I.}~\bibnamefont
  {Esteban}}, \bibinfo {author} {\bibfnamefont {M.~C.}\ \bibnamefont
  {Gonzalez-Garcia}}, \bibinfo {author} {\bibfnamefont {M.}~\bibnamefont
  {Maltoni}}, \bibinfo {author} {\bibfnamefont {T.}~\bibnamefont {Schwetz}},\
  and\ \bibinfo {author} {\bibfnamefont {A.}~\bibnamefont {Zhou}},\ }\bibfield
  {title} {\bibinfo {title} {{The fate of hints: updated global analysis of
  three-flavor neutrino oscillations}},\ }\href
  {https://doi.org/10.1007/JHEP09(2020)178} {\bibfield  {journal} {\bibinfo
  {journal} {JHEP}\ }\textbf {\bibinfo {volume} {09}},\ \bibinfo {pages}
  {178}},\ \Eprint {https://arxiv.org/abs/2007.14792} {arXiv:2007.14792
  [hep-ph]} \BibitemShut {NoStop}%
\bibitem [{\citenamefont {Ishihara}(2021)}]{Ishihara:2019aao}%
  \BibitemOpen
  \bibfield  {author} {\bibinfo {author} {\bibfnamefont {A.}~\bibnamefont
  {Ishihara}} (\bibinfo {collaboration} {IceCube}),\ }\bibfield  {title}
  {\bibinfo {title} {{The IceCube Upgrade - Design and Science Goals}},\ }\href
  {https://doi.org/10.22323/1.358.1031} {\bibfield  {journal} {\bibinfo
  {journal} {PoS}\ }\textbf {\bibinfo {volume} {ICRC2019}},\ \bibinfo {pages}
  {1031} (\bibinfo {year} {2021})},\ \Eprint {https://arxiv.org/abs/1908.09441}
  {arXiv:1908.09441 [astro-ph.HE]} \BibitemShut {NoStop}%
\bibitem [{\citenamefont {Abbasi}\ \emph
  {et~al.}(2025{\natexlab{d}})\citenamefont {Abbasi} \emph
  {et~al.}}]{IceCube:2025chb}%
  \BibitemOpen
  \bibfield  {author} {\bibinfo {author} {\bibfnamefont {R.}~\bibnamefont
  {Abbasi}} \emph {et~al.} (\bibinfo {collaboration} {IceCube}),\ }\bibfield
  {title} {\bibinfo {title} {{Physics potential of the IceCube Upgrade for
  atmospheric neutrino oscillations}},\ }\href@noop {} {\  (\bibinfo {year}
  {2025}{\natexlab{d}})},\ \Eprint {https://arxiv.org/abs/2509.13066}
  {arXiv:2509.13066 [hep-ex]} \BibitemShut {NoStop}%
\bibitem [{\citenamefont {Adrian-Martinez}\ \emph {et~al.}(2016)\citenamefont
  {Adrian-Martinez} \emph {et~al.}}]{KM3Net:2016zxf}%
  \BibitemOpen
  \bibfield  {author} {\bibinfo {author} {\bibfnamefont {S.}~\bibnamefont
  {Adrian-Martinez}} \emph {et~al.} (\bibinfo {collaboration} {KM3Net}),\
  }\bibfield  {title} {\bibinfo {title} {{Letter of intent for KM3NeT 2.0}},\
  }\href {https://doi.org/10.1088/0954-3899/43/8/084001} {\bibfield  {journal}
  {\bibinfo  {journal} {J. Phys. G}\ }\textbf {\bibinfo {volume} {43}},\
  \bibinfo {pages} {084001} (\bibinfo {year} {2016})},\ \Eprint
  {https://arxiv.org/abs/1601.07459} {arXiv:1601.07459 [astro-ph.IM]}
  \BibitemShut {NoStop}%
\bibitem [{\citenamefont {Abi}\ \emph {et~al.}(2021)\citenamefont {Abi} \emph
  {et~al.}}]{DUNE:2021cuw}%
  \BibitemOpen
  \bibfield  {author} {\bibinfo {author} {\bibfnamefont {B.}~\bibnamefont
  {Abi}} \emph {et~al.} (\bibinfo {collaboration} {DUNE}),\ }\bibfield  {title}
  {\bibinfo {title} {{Experiment Simulation Configurations Approximating DUNE
  TDR}},\ }\href@noop {} {\  (\bibinfo {year} {2021})},\ \Eprint
  {https://arxiv.org/abs/2103.04797} {arXiv:2103.04797 [hep-ex]} \BibitemShut
  {NoStop}%
\bibitem [{\citenamefont {Abe}\ \emph {et~al.}(2018)\citenamefont {Abe} \emph
  {et~al.}}]{Hyper-Kamiokande:2018ofw}%
  \BibitemOpen
  \bibfield  {author} {\bibinfo {author} {\bibfnamefont {K.}~\bibnamefont
  {Abe}} \emph {et~al.} (\bibinfo {collaboration} {Hyper-Kamiokande}),\
  }\bibfield  {title} {\bibinfo {title} {{Hyper-Kamiokande Design Report}},\
  }\href@noop {} {\  (\bibinfo {year} {2018})},\ \Eprint
  {https://arxiv.org/abs/1805.04163} {arXiv:1805.04163 [physics.ins-det]}
  \BibitemShut {NoStop}%
\bibitem [{\citenamefont {Eller}\ \emph {et~al.}(2023)\citenamefont {Eller}
  \emph {et~al.}}]{IceCube:2023ahv}%
  \BibitemOpen
  \bibfield  {author} {\bibinfo {author} {\bibfnamefont {P.}~\bibnamefont
  {Eller}} \emph {et~al.} (\bibinfo {collaboration} {IceCube}),\ }\bibfield
  {title} {\bibinfo {title} {{A model independent parametrization of the
  optical properties of the refrozen IceCube drill holes}},\ }\href
  {https://doi.org/10.22323/1.444.1034} {\bibfield  {journal} {\bibinfo
  {journal} {PoS}\ }\textbf {\bibinfo {volume} {ICRC2023}},\ \bibinfo {pages}
  {1034} (\bibinfo {year} {2023})},\ \Eprint {https://arxiv.org/abs/2307.15298}
  {arXiv:2307.15298 [astro-ph.HE]} \BibitemShut {NoStop}%
\bibitem [{\citenamefont {Barr}\ \emph {et~al.}(2006)\citenamefont {Barr},
  \citenamefont {Gaisser}, \citenamefont {Robbins},\ and\ \citenamefont
  {Stanev}}]{Barr:2006it}%
  \BibitemOpen
  \bibfield  {author} {\bibinfo {author} {\bibfnamefont {G.~D.}\ \bibnamefont
  {Barr}}, \bibinfo {author} {\bibfnamefont {T.~K.}\ \bibnamefont {Gaisser}},
  \bibinfo {author} {\bibfnamefont {S.}~\bibnamefont {Robbins}},\ and\ \bibinfo
  {author} {\bibfnamefont {T.}~\bibnamefont {Stanev}},\ }\bibfield  {title}
  {\bibinfo {title} {{Uncertainties in Atmospheric Neutrino Fluxes}},\ }\href
  {https://doi.org/10.1103/PhysRevD.74.094009} {\bibfield  {journal} {\bibinfo
  {journal} {Phys. Rev. D}\ }\textbf {\bibinfo {volume} {74}},\ \bibinfo
  {pages} {094009} (\bibinfo {year} {2006})},\ \Eprint
  {https://arxiv.org/abs/astro-ph/0611266} {arXiv:astro-ph/0611266}
  \BibitemShut {NoStop}%
\bibitem [{\citenamefont {Fedynitch}\ \emph {et~al.}(2019)\citenamefont
  {Fedynitch}, \citenamefont {Riehn}, \citenamefont {Engel}, \citenamefont
  {Gaisser},\ and\ \citenamefont {Stanev}}]{Fedynitch:2018cbl}%
  \BibitemOpen
  \bibfield  {author} {\bibinfo {author} {\bibfnamefont {A.}~\bibnamefont
  {Fedynitch}}, \bibinfo {author} {\bibfnamefont {F.}~\bibnamefont {Riehn}},
  \bibinfo {author} {\bibfnamefont {R.}~\bibnamefont {Engel}}, \bibinfo
  {author} {\bibfnamefont {T.~K.}\ \bibnamefont {Gaisser}},\ and\ \bibinfo
  {author} {\bibfnamefont {T.}~\bibnamefont {Stanev}},\ }\bibfield  {title}
  {\bibinfo {title} {{Hadronic interaction model sibyll 2.3c and inclusive
  lepton fluxes}},\ }\href {https://doi.org/10.1103/PhysRevD.100.103018}
  {\bibfield  {journal} {\bibinfo  {journal} {Phys. Rev. D}\ }\textbf {\bibinfo
  {volume} {100}},\ \bibinfo {pages} {103018} (\bibinfo {year} {2019})},\
  \Eprint {https://arxiv.org/abs/1806.04140} {arXiv:1806.04140 [hep-ph]}
  \BibitemShut {NoStop}%
\bibitem [{\citenamefont {Tena-Vidal}\ \emph {et~al.}(2021)\citenamefont
  {Tena-Vidal} \emph {et~al.}}]{GENIE-cross}%
  \BibitemOpen
  \bibfield  {author} {\bibinfo {author} {\bibfnamefont {J.}~\bibnamefont
  {Tena-Vidal}} \emph {et~al.} (\bibinfo {collaboration} {GENIE}),\ }\bibfield
  {title} {\bibinfo {title} {{Neutrino-nucleon cross-section model tuning in
  GENIE v3}},\ }\href {https://doi.org/10.1103/PhysRevD.104.072009} {\bibfield
  {journal} {\bibinfo  {journal} {Phys. Rev. D}\ }\textbf {\bibinfo {volume}
  {104}},\ \bibinfo {pages} {072009} (\bibinfo {year} {2021})},\ \Eprint
  {https://arxiv.org/abs/2104.09179} {arXiv:2104.09179 [hep-ph]} \BibitemShut
  {NoStop}%
\bibitem [{\citenamefont {Cooper-Sarkar}\ \emph {et~al.}(2011)\citenamefont
  {Cooper-Sarkar}, \citenamefont {Mertsch},\ and\ \citenamefont
  {Sarkar}}]{Cooper-Sarkar:2011jtt}%
  \BibitemOpen
  \bibfield  {author} {\bibinfo {author} {\bibfnamefont {A.}~\bibnamefont
  {Cooper-Sarkar}}, \bibinfo {author} {\bibfnamefont {P.}~\bibnamefont
  {Mertsch}},\ and\ \bibinfo {author} {\bibfnamefont {S.}~\bibnamefont
  {Sarkar}},\ }\bibfield  {title} {\bibinfo {title} {{The high energy neutrino
  cross-section in the Standard Model and its uncertainty}},\ }\href
  {https://doi.org/10.1007/JHEP08(2011)042} {\bibfield  {journal} {\bibinfo
  {journal} {JHEP}\ }\textbf {\bibinfo {volume} {08}},\ \bibinfo {pages}
  {042}},\ \Eprint {https://arxiv.org/abs/1106.3723} {arXiv:1106.3723 [hep-ph]}
  \BibitemShut {NoStop}%
\end{thebibliography}%

\onecolumngrid
\begin{center}
\rule{0.5\textwidth}{0.5pt}
\end{center}

%%%%%%%%%%%%%%%%%%%%%%%%%%%%%%%%%%%%%%%%%%%%%%%%%%%%%%%%%%%%%%%%%%%%%%%%%%%%%%%

\newpage
\clearpage

\appendix

\onecolumngrid

\begin{center}
	\large
	Supplemental Material for\\
	\smallskip
	{\it First Constraints on Long-Range Neutrino Interactions using IceCube DeepCore}
\end{center}

\twocolumngrid

\section{Appendix A: Systematic Uncertainties and their Best-fit Values}

Table~\ref{tab:systematic_params} summarizes all the systematic parameters considered in this analysis as nuisance parameters. During the fit to the data, the $\chi^2_{\rm mod}$ is minimized over a total of 20 nuisance parameters simultaneously, along with  the corresponding physics parameter, $V_{e\mu}$ (or $V_{e\tau}$). Each parameter is listed with its nominal value, a Gaussian prior of $1\sigma$ if available, and the fit range. A Gaussian pull penalty is added to $\chi^2_{\rm mod}$ for the parameters that have priors around the nominal values.
	
\onecolumngrid

\setcounter{table}{0}
\renewcommand{\thetable}{A\arabic{table}}
\begin{table}[h!]
	\centering
	\renewcommand{\arraystretch}{1.3}
	\begin{tabular}{l@{\hskip 15pt}c@{\hskip 15pt}c@{\hskip 15pt}c@{\hskip 15pt}c@{\hskip 15pt}c}
		\hline
		\hline
		\textbf{Parameter} & \textbf{Best-fit} ($V_{e\mu}$) & \textbf{Best-fit} ($V_{e\tau}$) &\textbf{Nominal value} & \textbf{Prior} & \textbf{Range} \\
		\hline
		\multicolumn{6}{@{}l}{\textbf{Detector:}} \\
		DOM efficiency		     & 1.064 	& 1.064 		& 1.0 		& $\pm\,$0.1 & [0.8, 1.2] \\
		Ice absorption           & 0.974 	& 0.974 		& 1.0 		& -&[0.9, 1.1] \\
		Ice scattering           & 0.988 	& 0.988 		& 1.05 		& -&[0.95, 1.15] \\
		Relative eff. $p_0$      & $-\,0.269$ 	& $-\,0.269$ 	& 0.10 		& -&[$-\,0.2$, 0.6] \\
		Relative eff. $p_1$      & $-\,0.043$ & $-\,0.043$ 	& $-\,0.05$ 	& -&[$-\,0.2$, 0.2] \\
		\hline
		\multicolumn{6}{@{}l}{\textbf{Atmospheric flux:}} \\
		$\Delta \gamma_{\nu}$     & 0.064 & 0.064 & 0.0 & $\pm\,$0.1 & [$-\,0.5$, 0.5] \\
		$\Delta \pi^\pm \text{ yields [A-F]}$ & 0.061 & 0.061 & 0.0 & $\pm\,$0.3 & [$-\,1.5$, 1.5] \\
		$\Delta \pi^\pm \text{ yields G}$& $-\,0.055$ & $-\,0.055$ & 0.0 & $\pm\,$0.15 & [$-\,1.5$, 1.5] \\
		$\Delta \pi^\pm \text{ yields H}$& $-\,0.018$ & $-\,0.018$ & 0.0 & $\pm\,$0.15 & [$-\,0.75$, 0.75] \\
		$\Delta K^+ \text{ yields W}$ & 0.085 & 0.085 & 0.0 & $\pm\,$0.4 & [$-\,2.0$, 2.0]\\
		$\Delta K^+ \text{ yields Y}$ & 0.107 & 0.108 & 0.0 & $\pm\,$0.3 & [$-\,1.5$, 1.5]\\
		$\Delta K^- \text{ yields W}$ & $-\,0.009$ & $-\,0.009$ & 0.0 & $\pm\,$0.4 & [$-\,2.0$, 2.0]\\
		\hline
		\multicolumn{6}{@{}l}{\textbf{Cross-section:}} \\
		$M_A^{\text{CCQE}}$ (in $\sigma$)  & 0.062 & 0.062  & 0.0 & $\pm\,$1.0 & [$-2.0$, 2.0]\\
		$M_A^{\text{CCRES}}$ (in $\sigma$)& 0.606 & 0.606 & 0.0 & $\pm\,$1.0 & [$-2.0$, 2.0]\\
		DIS CSMS                & 0.034 & 0.035 & 0.0 & $\pm\,$1.0 & [$-3.0$, 3.0]\\
		$\sigma_{\rm NC}/\sigma_{\rm CC} $ & 1.127 & 1.127 & 1.0 & $\pm\,$0.2 & [0.5, 1.5]\\
		\hline
		\multicolumn{6}{@{}l}{\textbf{Normalization:}} \\
		$A_{\text{eff}}$ scale   & 0.824 & 0.824 & 1.0 & -&[0.6, 1.4] \\
		\hline
		\multicolumn{6}{@{}l}{\textbf{Atmospheric muons:}} \\
		Atm. $\mu$ scale         & 1.365 & 1.365 & 1.0 & -&[0.7, 1.5] \\
		\hline
		\multicolumn{6}{@{}l}{\textbf{Oscillations:}} \\
		$\theta_{23}$           & 45.385$^\circ$ & 45.297$^\circ$ & 45.573$^\circ$ & -&[38$^\circ$, 52$^\circ$] \\
		$\Delta m^2_{31}$       & 0.002489 eV$^2$ & 0.002489 eV$^2$ & 0.002484 eV$^2$ & -&[0.002, 0.003] eV$^2$ \\
		\hline
		\hline
	\end{tabular}
	
	\caption{A consolidated list of the systematic uncertainty parameters considered in this analysis as nuisance parameters with their nominal values, $1\sigma$ priors (if available), and allowed ranges for fitting. The second and third columns list the best-fit values for the hypothesis with $L_e - L_{\mu}$ and $L_e - L_{\tau}$ symmetries, respectively, which are considered one at a time. We consider normal mass ordering while fitting.}
	\label{tab:systematic_params}
\end{table}
\twocolumngrid

We follow the treatment of systematic uncertainties as described in detail in Ref.~\cite{IceCubeCollaboration:2023wtb}. For uncertainties related to the detector response, a total of five parameters are considered. The photon detection efficiency of each Digital Optical Module (DOM) is parameterized by an overall DOM efficiency scale. The refrozen ice around the DOMs in holes has different properties from the original ice. The effect of refrozen ice on the angular acceptance of the module is modeled in terms of the two parameters $p_0$ and $p_1$~\cite{IceCube:2023ahv}. The optical properties of the ice medium are characterized using two scaling parameters to account for the scattering and absorption lengths.

The atmospheric neutrino flux predicted by Honda et al.~\cite{Honda:2015fha} is used as the baseline flux model for the calculations. The flux uncertainties related to the hadron production are incorporated following the prescription in Ref.~\cite{Barr:2006it} and are evaluated using MCEq~\cite{Fedynitch:2018cbl}. The systematic treatment incorporates the effective parameters describing uncertainties in kaon and pion production, as well as an uncertainty in the spectral index. In addition, overall normalization factors are included for neutrinos and atmospheric muons separately, ensuring that the oscillation measurement remains independent of the absolute flux predictions.

\setcounter{figure}{0}
\renewcommand{\thefigure}{A\arabic{figure}}
\begin{figure}[t!]   % capital H forces placement exactly here
	\centering
	\includegraphics[width=\linewidth]{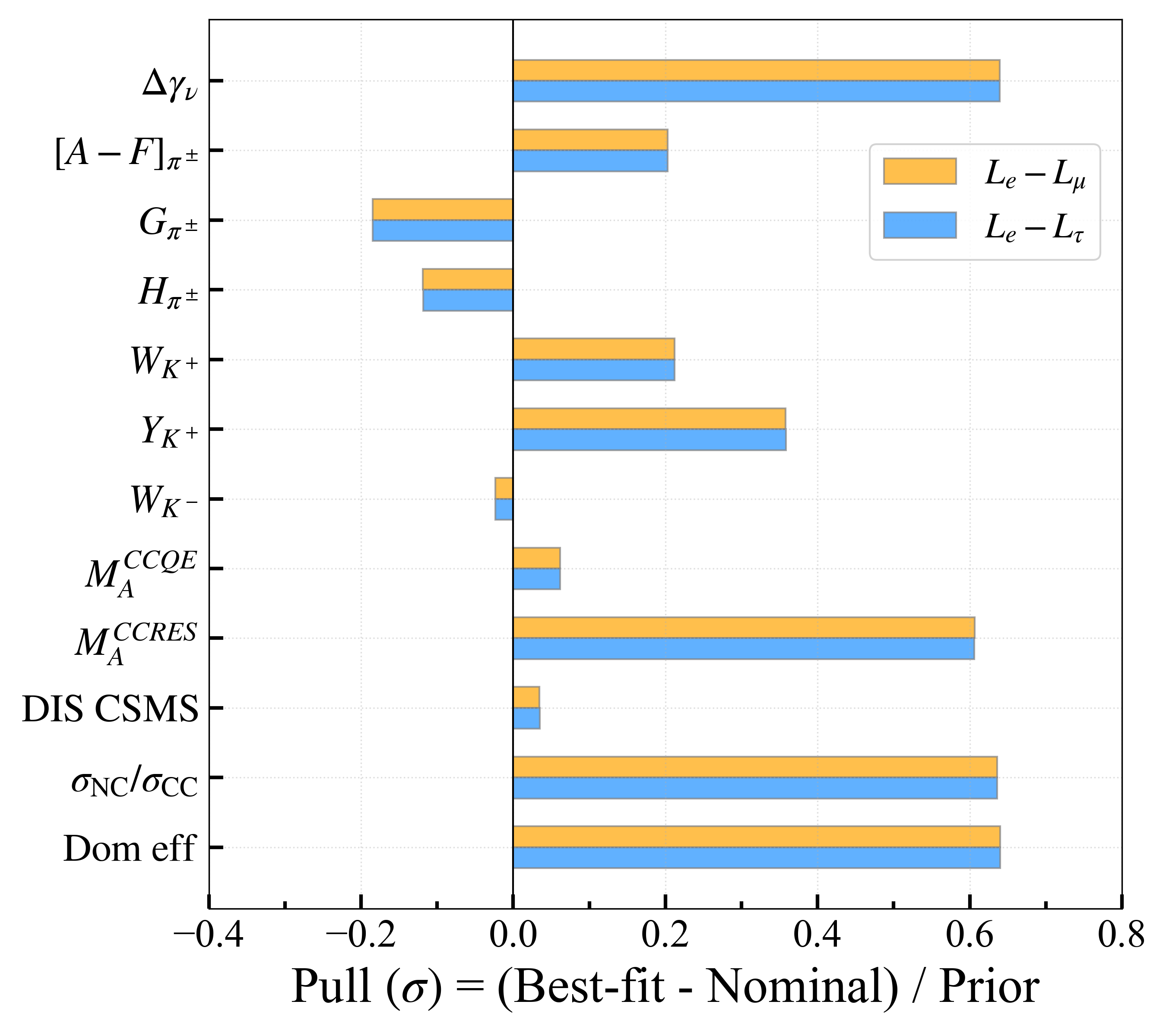}
	\caption{Pulls of systematic parameter values obtained from the data fitting for $L_e - L_{\mu}$ and $L_e - L_{\tau}$ symmetries considered one at a time. Here, only those systematic parameters are shown where prior is available.}
	\label{fig:param_pull}
\end{figure}
 
We use the neutrino event generator GENIE as the baseline model for the neutrino-nucleon interaction cross section. The re-weighting factors for the axial mass $M_A^{\rm CCQE}$ and $M_A^{\rm RES}$, corresponding to charged-current quasi-elastic and resonant interactions, respectively, modify the corresponding total cross sections. These two parameters are considered as nuisance parameters by varying them within their $2\sigma$ uncertainties. For the deep-inelastic scattering (DIS), an interpolated model between the predictions of GENIE~\cite{GENIE-cross} and CSMS~\cite{Cooper-Sarkar:2011jtt} is employed, which is described by a parameter called ``DIS CSMS''. A value of zero for DIS CSMS corresponds to the GENIE model, while a value of one corresponds to the CSMS model.

The best-fit values of these nuisance parameters, together with the two oscillation parameters, are listed in Tab.~\ref{tab:systematic_params}. The second and third columns correspond to the fits with $L_e - L_\mu$ and $L_e - L_\tau$ symmetries, respectively, which are considered one-at-a-time. Figure~\ref {fig:param_pull} shows the pulls for the nuisance parameters that have Gaussian priors. The orange and the blue color bars correspond to the pulls for the fits with the  $L_e - L_\mu$ and $L_e - L_\tau$ symmetries, respectively. From the figure, we can observe that all pulls remain well within $1\sigma$ range of their corresponding priors. Furthermore, the best-fit values of the remaining parameters, which do not have priors, lie well within their allowed fit ranges as summarized in Table~\ref{tab:systematic_params}.

\onecolumngrid
\setcounter{figure}{0}
\renewcommand{\thefigure}{B\arabic{figure}}
\begin{figure*}[htp!]
	\centering
	\includegraphics[width=\linewidth]{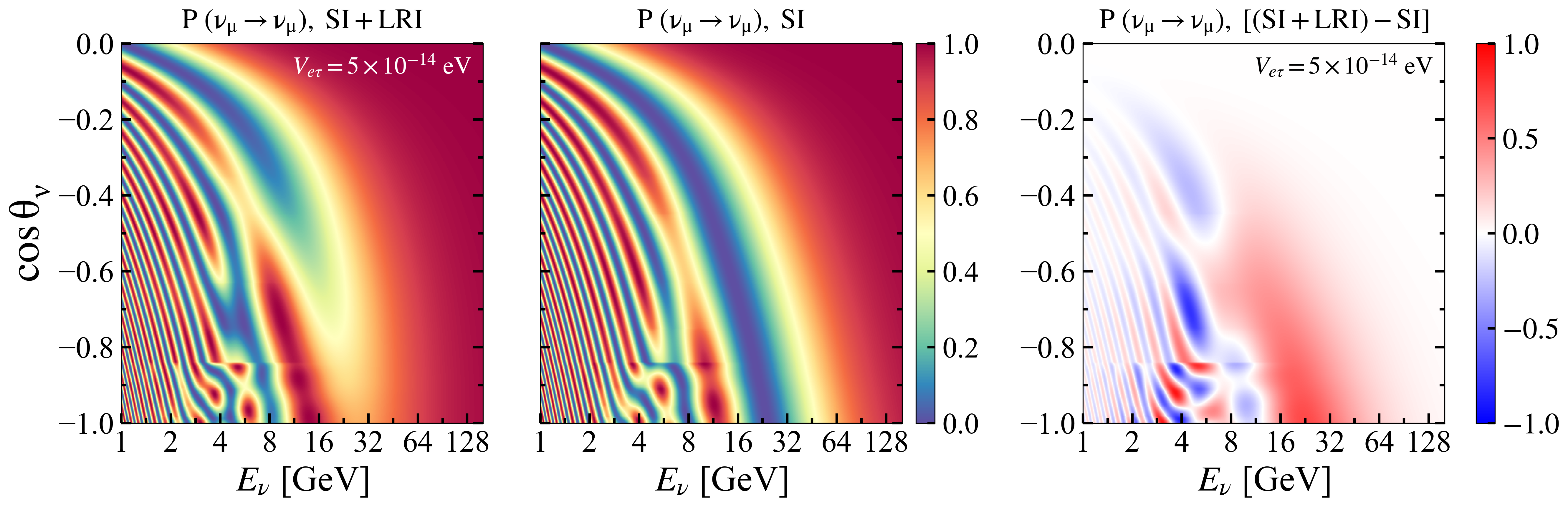}
	\caption{$P(\nu_\mu \rightarrow \nu_\mu)$ oscillograms shown in $(E_{\nu},\,\cos\theta_{\nu})$ plane for SI and SI + LRI ($V_{e\tau} = 5 \times 10^{-14}\ \text{eV}$) scenarios in the middle and the left panels, respectively, assuming NO. The right panel illustrates the difference between the SI + LRI and SI scenarios. Here, we consider $\theta_{23} = 45.57^\circ$ and $\Delta m^2_{31} = 2.48 \times 10^{-3}~{\rm eV}^2$.}
	\label{fig:prob_V_etau}
\end{figure*}
\clearpage
\twocolumngrid

\section{Appendix B: Effect of the $L_e - L_\tau$ Symmetry on Oscillograms and Events}
Similar to the $L_e - L_\mu$ case, the presence of a nonzero $V_{e\tau}$ leads to characteristic distortions in the oscillation pattern across the $(E_{\nu},\cos\theta_{\nu})$ plane. Figure~\ref{fig:prob_V_etau} shows the impact of the $L_e - L_\tau$ symmetry on $P(\nu_\mu \to \nu_\mu)$ oscillogram. The middle panel shows the case of the standard interactions (SI) without LRI, and the left panel corresponds to the SI + LRI scenario with $V_{e\tau} = 5 \times 10^{-14}$~eV. The right panel illustrates the difference between the SI + LRI and SI scenarios. We observe that the presence of $V_{e\tau}$ results in the disappearance of the oscillation valley (blue band) around energies of 16 GeV to 32 GeV and $\cos\theta < -0.7$. The same effect can be seen in the form of a prominent red patch in the probability difference oscillograms in the right panel. This feature can be interpreted from the approximate dependence of effective mixing angle $\theta^m_{23}$ on $V_{e\tau}$~\cite{Khatun:2018lzs} expressed as,
\begin{equation}
	\tan 2\theta^m_{23} = \frac{\cos^{2}\theta_{13}
		- \alpha \cos^{2}\theta_{12}
		+ \alpha \sin^{2}\theta_{12} \sin^{2}\theta_{13}}
	{-\,W + \alpha \sin 2\theta_{12} \sin\theta_{13}},
\end{equation}
where $W \equiv 2EV_{e\tau}/\Delta m^2_{31}$ and $\alpha \equiv \Delta m^2_{21}/\Delta m^2_{31}$. In this expression, an increase in $V_{e\tau}$ drives $\theta^m_{23}$
to larger values, thereby causing the oscillation valley to vanish, as seen in Fig.~\ref{fig:prob_V_etau}.

\begin{figure}[htp!]
	\centering
	\includegraphics[width=\linewidth]{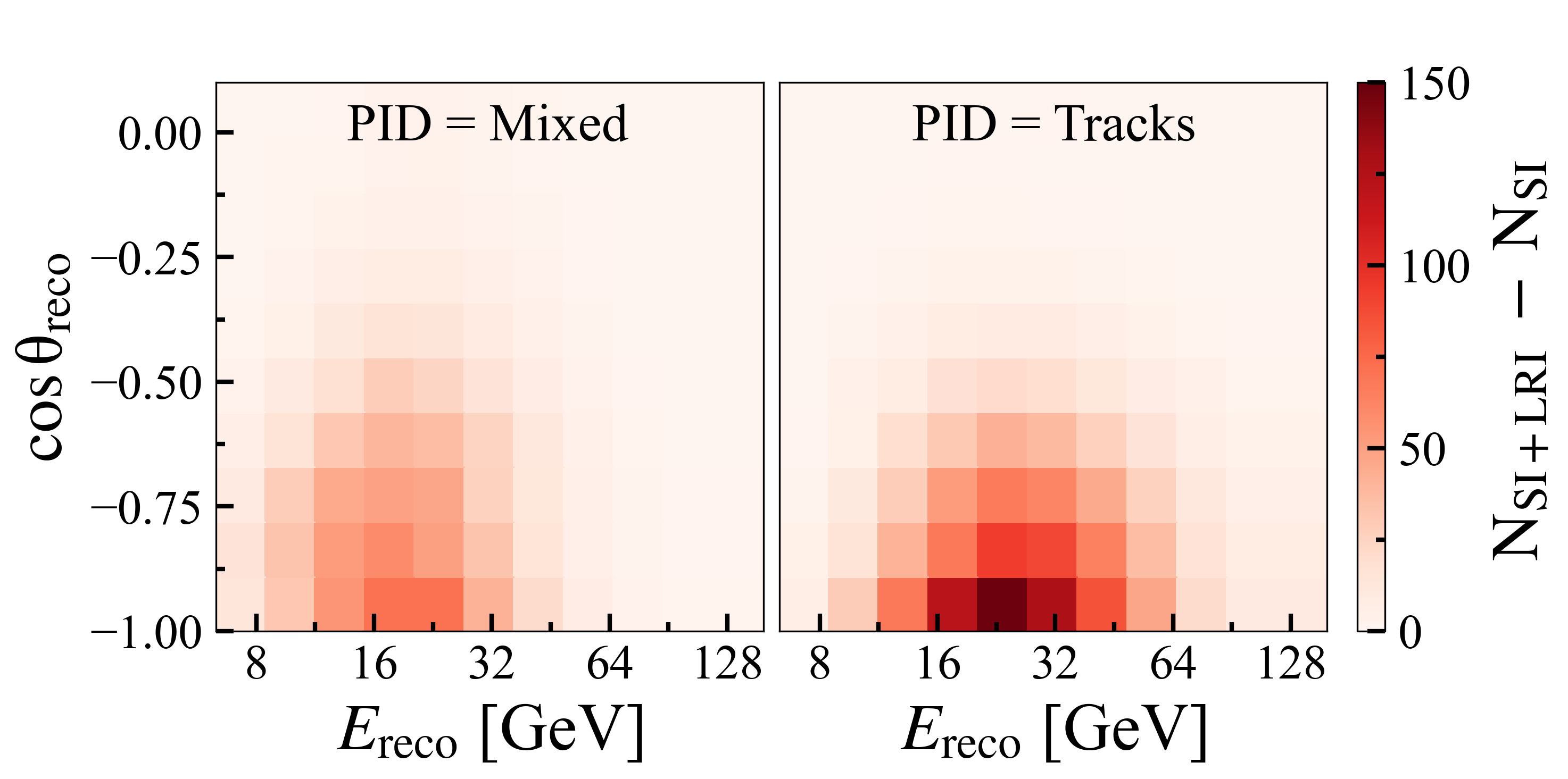}
	\caption{The difference of expected MC events at DeepCore for SI + LRI ($V_{e\tau} = 5 \times 10^{-14}$ eV) and SI scenarios. Here, we use the nominal values of the nuisance parameters as given in Table~\ref{tab:systematic_params}.}
	\label{fig:etau_event_diff}
\end{figure} 

Figure~\ref{fig:etau_event_diff} presents the effect of the $L_e - L_\tau$ symmetry on the expected MC events with the nominal values of the nuisance parameters. The distributions correspond to the difference of events for the SI + LRI ($V_{e\tau} = 5 \times 10^{-14}\ \text{eV}$) and SI scenarios. The left and right panels show the distributions for mixed and track-like events, respectively. We can observe that the effect of the presence of $V_{e\tau}$  is most prominent in the longer baseline and intermediate energy bins, which is consistent with the effects on oscillation probabilities as shown in Fig.~\ref{fig:prob_V_etau}.

\section{Appendix C: Goodness of fit and Data - MC Agreement }

The goodness of fit is evaluated by fitting the 500 pseudo-experiments, which are obtained by introducing statistical fluctuations to the expected number of events in every analysis bin of the Asimov template. We use the best-fit values of nuisance and physics parameters to produce the Asimov events. The pseudo-trials are fitted with $L_e - L_\mu$ and $L_e - L_\tau$ symmetries considered one at a time. The resulting $\chi^2_{\rm mod}$ distribution is shown in Fig.~\ref{fig:Goodness of Fit} for $L_e - L_\mu$ and $L_e - L_\tau$ symmetries with blue and red curves, respectively. The observed $\chi^2_{\rm mod}$ shown by vertical lines are used to calculate p-values, which correspond to the fraction of pseudo-trials with expected $\chi^2_{\rm mod}$ more than the observed $\chi^2_{\rm mod}$. The p-values for $L_e - L_\mu$ and $L_e - L_\tau$ symmetries are found to be 0.26 and 0.25, respectively, which indicate a good agreement between these models and the data.

\setcounter{figure}{0}
\renewcommand{\thefigure}{C\arabic{figure}}
\begin{figure}[htp!]
	\centering
	\includegraphics[width=\linewidth]{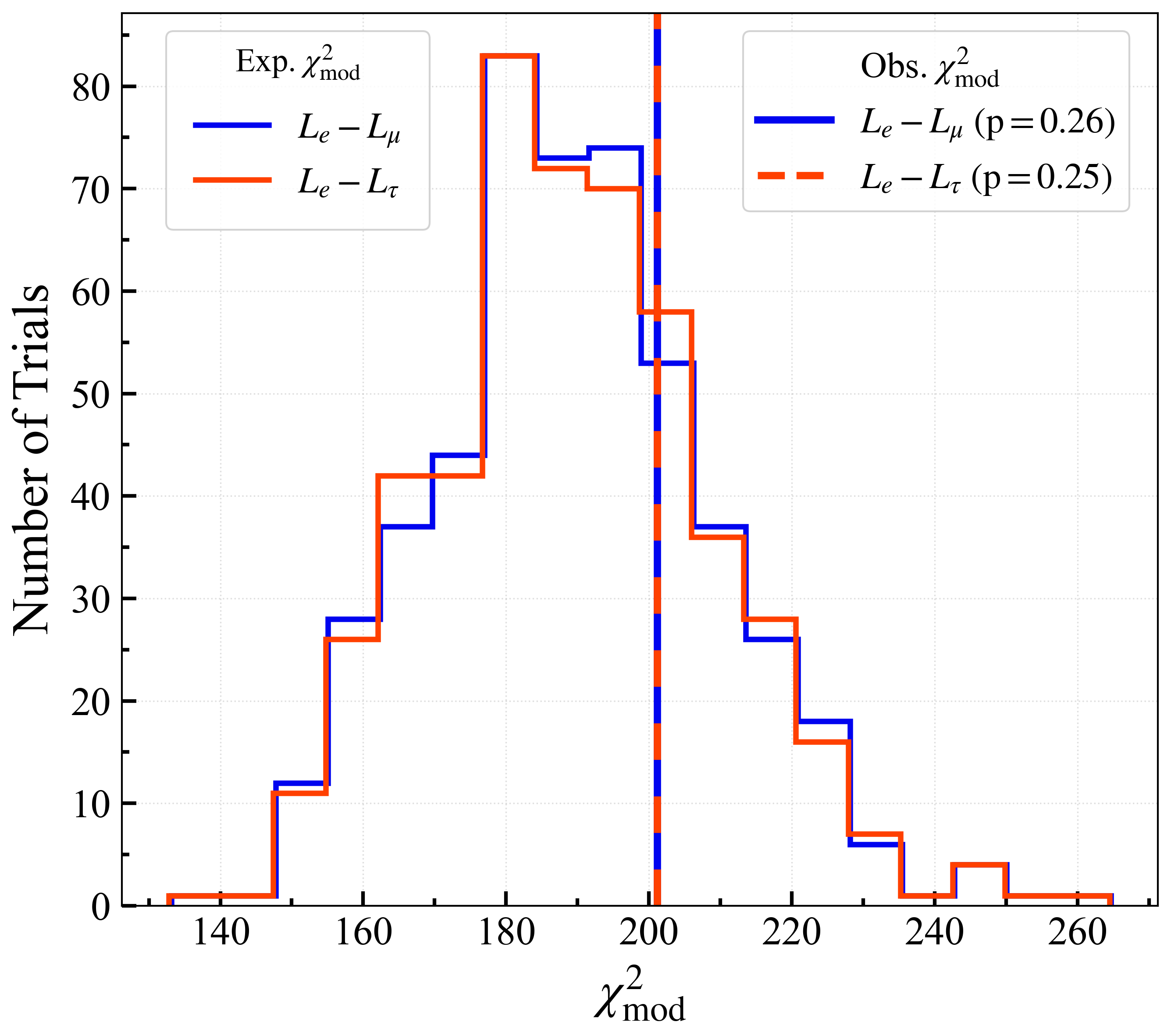}
	\caption{A comparison of the observed $\chi^2_{\rm mod}$ (vertical line) with the expected distribution of $\chi^2_{\rm mod}$ obtained by fitting the statistically fluctuated 500 pseudo-experiments. The blue and red curves correspond to $L_e - L_{\mu}$ and $L_e - L_{\tau}$ symmetries, respectively, which are considered one at a time. The goodness of fit in terms of the p-values are found to be 0.26 and 0.25 for $L_e - L_{\mu}$ and $L_e - L_{\tau}$ symmetries, respectively.}
	\label{fig:Goodness of Fit}
\end{figure} 

\begin{figure*}[t]
	\centering
	\begin{minipage}[b]{0.49\textwidth}
		\includegraphics[width=\linewidth]{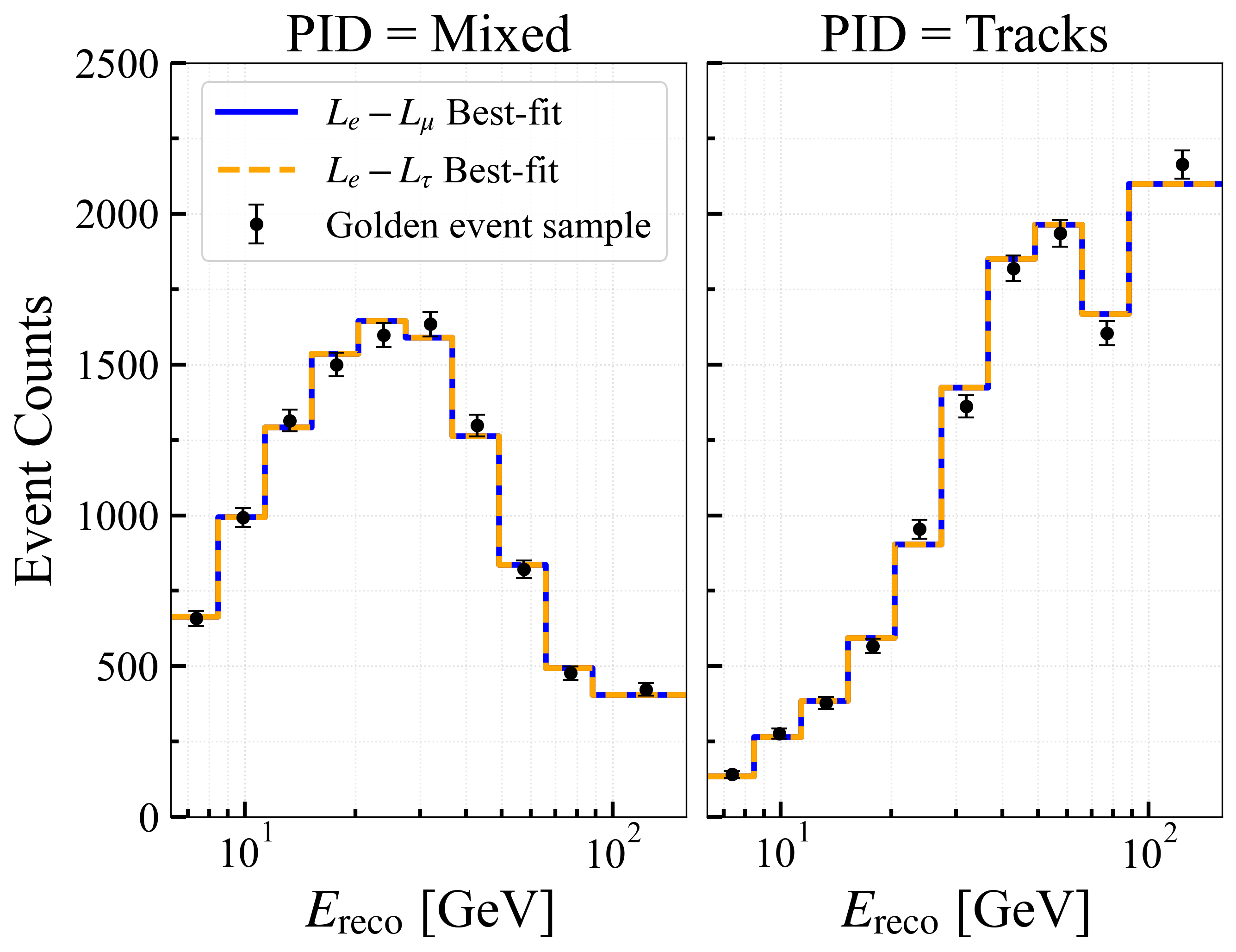}
	\end{minipage}
	\hfill
	\begin{minipage}[b]{0.49\textwidth}
		\includegraphics[width=\linewidth]{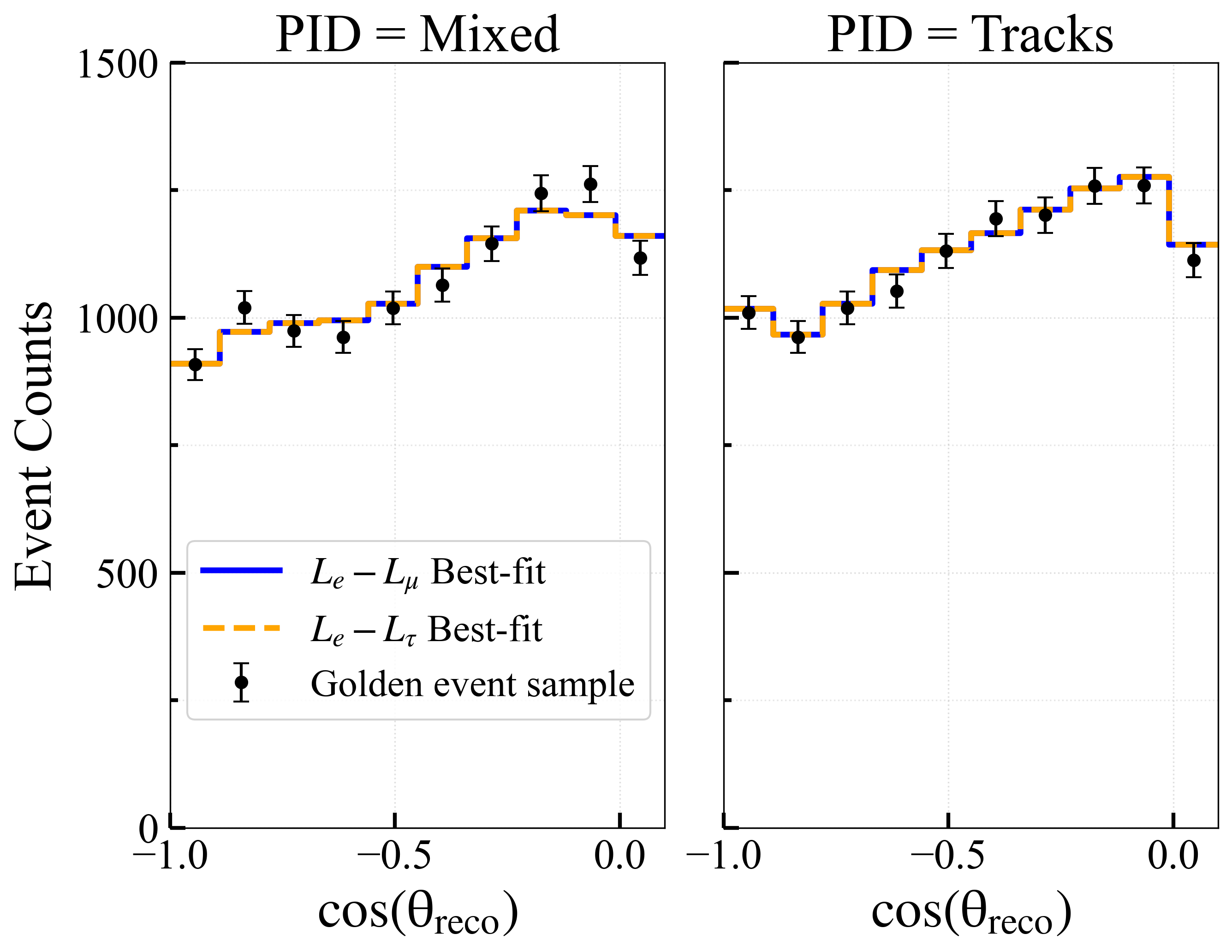}
	\end{minipage}
	\caption{Distributions of reconstructed energy and cosine of the zenith angle comparing the observed data with the best-fit MC for the 8-year golden event sample of IceCube DeepCore. The solid-blue and dashed-orange curves show the best-fit MC for hypotheses with $L_e -L_\mu$ and $L_e - L_\tau$ symmetries, respectively, considered one at a time. The distributions are presented separately for mixed and track-like events in the left and the right panels, respectively.}
	\label{fig:Data_MC}
\end{figure*}

The left and right panels of Fig.~\ref{fig:Data_MC} show the distributions of reconstructed neutrino energy and reconstructed cosine of zenith angle, respectively, comparing the observed data with the best-fit MC simulations for the 8-year golden event sample of IceCube DeepCore. Within each panel, the left and right sub-figures correspond to mixed and track-like event samples. The black dots with error bars correspond to the distribution of the observed data. The solid-blue and dashed-orange distributions denote the best-fit MC predictions for $L_e -L_\mu$ and $L_e - L_\tau$ symmetries, respectively, considered one at a time. We observe that, for both reconstructed observables, the best-fit predictions closely reproduce the observed data, demonstrating an overall good agreement between the models and the data.

\end{document}